\documentclass[usenatbib]{mn2e}
\usepackage{graphicx,fixltx2e}

\def\mrpd{\hbox{mrad\,d$^{-1}$}}
\def\gpcps{\hbox{g\,cm$^{-1}$\,s$^{-2}$}}
\def\chisq{\mbox{$\chi^2$}}
\def\chinu{\hbox{$\chi^2_\nu$}}
\def\msun{\hbox{${\rm M}_{\odot}$}}
\def\mspy{\hbox{${\rm M}_{\odot}$\,yr$^{-1}$}}
\def\rsun{\hbox{${\rm R}_{\odot}$}}

\def\rstar{\hbox{$R_{\star}$}}
\def\teff{\hbox{$T_{\rm eff}$}}
\def\logg{\hbox{$\log g$}}
\def\sn{\hbox{S/N}}

\def\kms{\hbox{km\,s$^{-1}$}}

\def\vesini{\hbox{$v_{\rm e}\sin(i)$}}
\def\pcc{\hbox{cm$^{-3}$}}

\def\ptt{\hbox{$10^{-4} I_{\rm c}$}}

\def\Bl{\hbox{$B_{\rm \ell}$}}

\def\degr{\hbox{$^\circ$}}

\def\vinf{\hbox{$v_{\infty}$}}
\def\Mdot{\hbox{$\dot{M}$}}
\def\ve{\hbox{$v_{\rm e}$}}

\newcommand{\civ}{C$\;${\sc iv}}
\newcommand{\nv}{N$\;${\sc v}}
\newcommand{\siiii}{Si$\;${\sc iii}}
\newcommand{\siiv}{Si$\;${\sc iv}}
\newcommand{\mgxi}{Mg$\;${\sc xi}}
\newcommand{\sixiii}{Si$\;${\sc xiii}}

\begin{document}

\title[The surprising magnetic topology of $\tau$~Sco] {The surprising
magnetic topology of $\tau$~Sco: fossil remnant or dynamo
output?\thanks{Based on observations obtained at the
Canada--France--Hawaii Telescope (CFHT) which is operated by the
National Research Council of Canada, the Institut National des Science
de l'Univers of the Centre National de la Recherche Scientifique of
France, and the University of Hawaii.} }

\makeatletter

\def\newauthor{%
  \end{author@tabular}\par
  \begin{author@tabular}[t]{@{}l@{}}}
\makeatother
 
\author[J.-F.~Donati et al.]  {\vspace{1.5mm}
J.-F.~Donati$^1$\thanks{E-mail: donati@ast.obs-mip.fr (J-FD);
idh@star.ucl.ac.uk (IDH); mmj@st-andrews.ac.uk (MMJ);
petit@ast.obs-mip.fr (PP); claude.catala@obspm.fr (CC); 
jlandstr@uwo.ca (JDL); jean-claude.bouret@oamp.fr (J-CB); 
evelyne.alecian@obspm.fr (EA); jrb3@st-andrews.ac.uk (JRB); 
forveill@cfht.hawaii.edu (TF); fpaletou@ast.obs-mip.fr (FP); 
manset@cfht.hawaii.edu (NM)}, 
I.D. Howarth$^2$, M.M. Jardine$^3$, P.~Petit$^1$, C. Catala$^4$, \\
\vspace{1.7mm}
{\hspace{-1.5mm}\LARGE\rm J.D. Landstreet$^5$, J.-C. Bouret$^6$, E.~Alecian$^4$, J.R.~Barnes$^3$, T.~Forveille$^7$,} \\ 
\vspace{1.5mm}
{\hspace{-1.5mm}\LARGE\rm F. Paletou$^1$ and N. Manset$^7$} \\
$^1$ LATT, Observatoire Midi-Pyr\'en\'ees, 14 Av.\ E.~Belin, F--31400 Toulouse, France \\
$^2$ Department of Physics and Astronomy, University College London,  Gower Street, London WC1E6BT, UK\\
$^3$ School of Physics and Astronomy, University of St~Andrews, St~Andrews, Scotland KY16 9SS, UK \\
$^4$ LESIA, CNRS--UMR 8109, Obs.\ de Paris, 5 Place Janssen, F--92195 Meudon Cedex, France \\ 
$^5$ Department of Physics and Astronomy, University of Western Ontario,  London Ontario N6A3K7, Canada\\ 
$^6$ LAM, Observatoire de Marseille-Provence, Traverse du Siphon BP 8, F--13376 Marseille Cedex 12, France\\
$^7$ CFHT, 65-1238 Mamalahoa Hwy, Kamuela HI, 96743 USA
}

\date{2006, MNRAS, submitted}
\maketitle
 
\begin{abstract} 
We report the discovery of a medium-strength ($\sim$0.5~kG) magnetic
field on the young, massive star $\tau$~Sco (B0.2$\;$V), which becomes
the third-hottest magnetic star known. Circularly polarized Zeeman
signatures are clearly detected in observations collected mostly with
the ESPaDOnS spectro{\-}polarimeter, recently installed on the 3.6-m
Canada--France--Hawaii Telescope; temporal variability is also clearly
established in the polarimetry, and can be unambiguously attributed to
rotational modulation with a period close to 41~d.  Archival UV
spectra confirm that this modulation repeats over timescales of
decades, and refine the rotation period to $41.033\pm0.002$~d.

Despite the slow rotation rate of $\tau$~Sco, we nonetheless succeed
in reconstructing the large-scale structure of its magnetic topology.
We find that the magnetic structure is unusually complex for a hot
star, with significant power in spherical-harmonic modes of degree up
to 5.  The surface topology is dominated by a potential field,
although a moderate toroidal component is probably present.  We fail
to detect {\em intrinsic} temporal variability of the magnetic
structure over the 1.5-yr period of our spectro{\-}polarimetric
observations (in agreement with the stable temporal variations of the
UV spectra), and infer that any differential surface rotation must be
very small.

The topology of the extended magnetic field that we derive from the
photo{\-}spheric magnetic maps is also more complex than a global 
dipole, and features in particular a significantly warped torus of closed 
magnetic loops encircling the star (tilted at about 90\degr\ to the 
rotation axis), with additional, smaller, networks of closed field lines.  
This topology appears to be consistent with the
exceptional X-ray properties of $\tau$~Sco and also provides a natural
explanation of the variability observed in wind-formed UV lines.
Although we cannot completely rule out the possibility that the field
is produced through dynamo processes of an exotic kind, we conclude 
that its magnetic field is most probably a fossil remnant from the 
star-formation stage.
\end{abstract}


\section{Introduction} 

Magnetic fields in hot, high-mass stars of spectral types O and
early~B may have a drastic influence on the physics of the stellar
interiors \citep[e.g.,][]{Spruit02} and atmospheres
\citep[e.g.,][]{Babel97}.  As a consequence, they can also
significantly modify these stars' long-term evolution, and in
particular their rotational history \citep[e.g.,][]{Maeder03,
Maeder04, Maeder05}.  However, quantifying these effects requires that
we know the basic properties of such fields.

From a theoretical point of view, several quite different mechanisms
have been proposed as potential means of generating large-scale
magnetic fields in very hot stars.  For example, such fields may be
fossil remnants of the star-formation stage, either as relics of the
field that pervaded the inter{\-}stellar medium from which the star
formed, or as left{\-}overs from dynamo action in the convective
Hayashi phase.  This hypothesis was initially proposed for the
lower-mass, chemically peculiar, magnetic Ap and Bp stars
\citep[e.g.,][]{Moss01} and has recently been rediscussed for the
particular case of high-mass O and early-B stars by \citet{Ferrario05,
Ferrario06}; in this scenario, very hot magnetic stars are expected to
be the progenitors of magnetic neutron stars.

A second possibility is that fields may be produced by continuing
dynamo action; this option, initially suggested about two decades ago
\citep[e.g.,][]{Moss82}, has not been closely examined until quite
recently, but is now attracting increasing interest from
theoreticians.  A dynamo action could operate in the convective core
\citep{Charbonneau01, Brun05}, but the outer radiative envelope could
also be involved in genuine dynamo processes, in a subsurface
shear layer \citep{Tout95, Lignieres96} or even throughout the whole
envelope \citep{Spruit02, Macdonald04, Mullan05, Maeder05, Braithwaite06}.

Unfortunately, all available theoretical options still suffer
significant problems.  Fossil-field theories have yet to demonstrate
that sufficient magnetic flux can survive the accelerated decay and
expulsion associated with Hayashi convection\footnote{This difficulty may not
be as problematical as it first seemed, as discussed in
\citet{Moss03}, and should not concern stars with masses larger than
10~\msun.}, while core-dynamo theories are still lacking detailed
models linking the field produced in the core with that emerging at the
surface \citep{Moss01}.  Being based on a very new idea,
radiative-zone dynamo theories are still in their infancy and need to
establish their potential validity, by demonstrating that, in
particular, stellar radiative zones are capable of sustaining
differential rotation throughout their whole volume.

In addition to these issues, existing theories face a number of
problems when compared to observations.  At some point, all theories
involve the coexistence of large-scale toroidal fields in stellar
interiors and possibly even close to the 
surface \citep[e.g.,][]{BraithSpruit04, BraithNord06}, which
observations of magnetic Ap and Bp stars do not yet confirm
\citep{Moss01}.  Moreover, dynamo theories predict magnetic topologies
that are expected to vary on short timescales and to depend strongly
on stellar rotation rates, which again is not observed in
intermediate-mass magnetic stars \citep{Moss01}.  Last but not least, 
dynamos theories should in principle apply to most hot stars (as they do 
for most cool stars), making it hard to understand why only a small 
fraction of them (about 10\% in the case of Ap/Bp stars) apparently hosts 
magnetic fields.  

Indirect evidence for the presence of magnetic fields in high-mass
stars is regularly reported in the literature, these being postulated as a
potential explanation for many otherwise enigmatic phenomena,
including unanticipated X-ray line profiles and high X-ray temperatures 
\citep[e.g.,][]{Cohen97,
Robinson02, Cohen03, Smith04}.  However, with {\em direct} detections
of magnetic fields in only two O stars to date \citep{Donati02,
Donati06}, together with less than a handful of early-B stars with
masses larger than 10~\msun\ \citep[e.g.,][]{Donati01, Neiner03}, very
little is known reliably about magnetic strengths and topologies from
a statistical point of view.  One reason for this is that absorption
lines of these stars are not only relatively few in number in the optical,
but are also generally rather broad (due to rotation, or to some other 
type of as yet unknown macroscopic mechanism; e.g., \citealt{Howarth97}),
decreasing dramatically the size of the Zeeman signatures that their
putative fields can induce.

In these respects, $\tau$~Sco (HR~6165, HD~149438, HIP~81266) appears
as a very promising target for magnetic-field studies; thanks to its
brightness and unusually narrow absorption lines (among the sharpest
known for stars more massive than 10~\msun), $\tau$~Sco is an obvious
candidate for accurate spectro{\-}polarimetric experiments.  
Moreover, its strong, hard X-ray emission
(with $\log L_{\rm X}/L_{\rm Bol}\simeq-6.5$, \citealt{Cohen97}) 
poses a severe challenge to the standard picture of O-star wind-shock
models, leading some authors \citep[e.g.,][]{Cohen03} to speculate
that it displays the presence of a magnetically confined wind, as
proposed by \citet{Stahl96} and \cite{Babel97}, and as detected on the
very young O~star $\theta^1$~Ori~C \citep{Donati02, Gagne05, Gagne05err}.  We therefore
included $\tau$~Sco in a list of candidates for observation with
ESPaDOnS, a high-efficiency spectro{\-}polarimeter installed in 2004/5
on the 3.6-m Canada--France--Hawaii Telescope (CFHT;
Donati et al., 2006, in preparation).

In this paper, we first present our new spectro{\-}polarimetric
observations and Zeeman detections obtained for $\tau$~Sco, and
briefly summarize the UV spectra (Sec.~\ref{sec:obs}).  We then carry
out a basic analysis of these data to establish the rotation period
(Sec.~\ref{sec:rot}), and review the fundamental stellar parameters in
the light of our result (Sec.~\ref{sec:params}).  We perform
detailed modelling of the magnetic topology of $\tau$~Sco, by direct
fitting to the observed Zeeman signatures (Sec.~\ref{sec:mod}).  Finally,
we discuss the implications of our results for models of the X-ray
emitting magneto{\-}sphere of $\tau$~Sco, as well for as theories of
large-scale magnetic-field generation in high-mass stars
(Secs.~\ref{sec:disc} and \ref{sec:disc2}).

\begin{table*}
\caption[]{Journal of spectropolarimetric observations.  Columns 1--6 
list the UT date \& time,   heliocentric Julian date 
(all at mid-exposure), observing site,  exposure time, and peak
signal to noise ratio (per 2.6~\kms\ velocity bin) of each
observation.  Columns 7 and 8 list the rms noise level (per 1.8~\kms\
velocity bin, relative to
the unpolarized continuum level $I_{\rm c}$) in
the circular-polarization profile produced by Least-Squares
Deconvolution (see Sec.~\ref{sec:obs}) and
the estimated longitudinal field \Bl\ (with 
corresponding 1$\sigma$ error bars). The rotational cycle $E$ 
from the ephemeris of eqn.~\ref{eq:ephem} is given in column~9.}  
\begin{tabular}{lccclrcrr}
\hline
\multicolumn{2}{c}{UT}                & HJD          & Obs  & 
\multicolumn{1}{c}{$t_{\rm exp}$} & 
\multicolumn{1}{c}{\sn}  & $\sigma_{\rm LSD}$ & 
\multicolumn{1}{c}{\Bl} & 
\multicolumn{1}{c}{Cycle} \\
\multicolumn{1}{c}{Date}   & (h:m:s)  & (2,453,000+) &      &   
\multicolumn{1}{c}{(s)}         &       &   (\ptt)  & 
\multicolumn{1}{c}{(G)} &  \\
\hline	      		 	       		       
2004 Sep.\ 04 & 05:58:31 & 252.7475     & CFHT & $4\times30$  & 1440 & 0.70 & $-24.4\pm3.3$ & 1.456 \\ 
2004 Sep.\ 04 & 06:23:17 & 252.7647     &      & $4\times30$  & 1280 & 0.75 & $-18.3\pm3.4$ & 1.457 \\ 
\hline	      		 	       		       
2004 Sep.\ 25 & 05:03:15 & 273.7075     & CFHT & $4\times30$  & 1060 & 0.96 & $+36.2\pm4.5$ & 1.967 \\ 
2004 Sep.\ 25 & 05:10:24 & 273.7124     &      & $2\times60$  & 1100 & 0.92 & $+47.5\pm4.3$ & 1.967 \\ 
\hline	      		 	       		       
2004 Sep.\ 26 & 08:51:15 & 274.8689     & AAT  & $4\times120$ &  890 & 1.03 & $+64.0\pm4.2$ & 1.995 \\ 
2004 Sep.\ 27 & 08:52:49 & 275.8700     &      & $4\times120$ &  860 & 1.10 & $+78.4\pm4.5$ & 2.020 \\ 
2004 Sep.\ 28 & 08:58:25 & 276.8739     &      & $4\times120$ &  670 & 1.36 & $+80.5\pm5.6$ & 2.044 \\ 
\hline	      		 	       		       
2005 May 23   & 09:14:25 & 513.8903     & CFHT & $4\times300$ & 1700 & 0.63 & $-47.9\pm3.0$ & 7.820 \\ 
2005 May 24   & 08:32:52 & 514.8614     &      & $4\times60$  &  900 & 1.17 & $-50.3\pm5.6$ & 7.844 \\ 
2005 May 25   & 08:41:35 & 515.8675     &      & $4\times60$  &  880 & 1.19 & $-41.4\pm5.5$ & 7.868 \\ 
\hline	      		 	       		       
2005 June  19 & 07:19:19 & 540.8096     & CFHT & $4\times120$ & 1200 & 0.85 & $-19.2\pm4.3$ & 8.476 \\ 
2005 June  20 & 06:44:15 & 541.7852     &      & $4\times120$ & 1550 & 0.66 & $-20.2\pm3.3$ & 8.500 \\ 
2005 June  21 & 06:51:16 & 542.7900     &      & $4\times120$ & 1650 & 0.62 & $-19.8\pm2.9$ & 8.525 \\ 
2005 June  22 & 08:42:17 & 543.8671     &      & $4\times120$ & 1440 & 0.71 & $-12.8\pm3.6$ & 8.551 \\ 
2005 June  23 & 06:07:24 & 544.7595     &      & $4\times120$ & 1670 & 0.61 & $-5.9\pm3.1$ & 8.572 \\ 
2005 June  24 & 06:22:32 & 545.7699     &      & $4\times120$ & 1710 & 0.60 & $+1.3\pm3.1$ & 8.597 \\ 
2005 June  25 & 06:05:50 & 546.7583     &      & $4\times120$ & 1570 & 0.68 & $+4.7\pm3.4$ & 8.621 \\ 
2005 June  26 & 06:00:31 & 547.7545     &      & $4\times120$ & 1590 & 0.66 & $+0.9\pm3.3$ & 8.646 \\ 
2005 June  26 & 11:02:47 & 547.9644     &      & $4\times120$ & 1370 & 0.82 & $-5.1\pm4.2$ & 8.651 \\ 
\hline	      		 	       		       
2005 Aug.\ 19 & 05:20:15 & 601.7224     & CFHT & $4\times30$  & 1250 & 0.77 & $+44.5\pm3.7$ & 9.961 \\ 
2005 Aug.\ 21 & 05:21:43 & 603.7232     &      & $4\times30$  & 1200 & 0.81 & $+80.3\pm4.0$ & 10.010 \\ 
2005 Aug.\ 23 & 05:11:45 & 605.7161     &      & $4\times30$  & 1150 & 0.85 & $+87.8\pm4.2$ & 10.058 \\ 
\hline	      		 	       		       
2005 Sep.\ 19 & 05:03:10 & 632.7079     & CFHT & $4\times30$  & 1180 & 0.83 & $-21.8\pm3.7$ & 10.716 \\ 
2005 Sep.\ 20 & 05:01:56 & 633.7070     &      & $4\times30$  & 1280 & 0.76 & $-28.9\pm3.6$ & 10.740 \\ 
2005 Sep.\ 24 & 04:55:44 & 637.7023     &      & $2\times50$  &  680 & 1.48 & $-45.9\pm6.8$ & 10.838 \\ 
2005 Sep.\ 25 & 04:48:19 & 638.6971     &      & $4\times30$  & 1200 & 0.83 & $-51.9\pm5.3$ & 10.862 \\ 
2005 Sep.\ 25 & 05:01:33 & 638.7063     &      & $4\times30$  & 1120 & 0.89 & $-43.4\pm4.2$ & 10.862 \\ 
\hline
2006 Feb.\ 07 & 15:03:44 & 774.1269    & CFHT & $4\times30$  & 1090 & 0.88 & $+44.9\pm4.1$ & 14.162 \\ 
2006 Feb.\ 08 & 16:22:33 & 775.1817    &      & $4\times30$  &  910 & 1.04 & $+44.3\pm4.8$ & 14.188 \\ 
2006 Feb.\ 09 & 16:16:56 & 776.1779    &      & $4\times30$  &  620 & 1.58 & $+38.9\pm7.0$ & 14.212 \\ 
2006 Feb.\ 10 & 16:33:25 & 777.1895    &      & $4\times30$  & 1250 & 0.76 & $+32.8\pm3.4$ & 14.237 \\ 
2006 Feb.\ 11 & 16:18:20 & 778.1791    &      & $4\times30$  & 1020 & 0.94 & $+25.8\pm4.2$ & 14.261 \\ 
2006 Feb.\ 13 & 16:15:23 & 780.1772    &      & $4\times30$  &  490 & 2.06 & $+35.4\pm9.4$ & 14.310 \\ 
2006 Feb.\ 14 & 16:23:12 & 781.1827    &      & $4\times30$  &  840 & 1.13 & $+0.6\pm5.6$ & 14.334 \\ 
2006 Feb.\ 15 & 14:07:12 & 782.0884    &      & $4\times30$  & 1060 & 0.92 & $+8.0\pm4.5$ & 14.357 \\ 
\hline
\end{tabular}
\label{tab:log}
\end{table*}

\section{Observations}
\label{sec:obs}

\subsection{Optical spectropolarimetry}

Spectropolarimetric observations of $\tau$~Sco were collected with
ESPaDOnS from 2004~Sep.\ to 2006~Feb., during the first engineering
runs (in 2004, when the field was first detected) and subsequently for
scheduled ESPaDOnS programmes.  The ESPaDOnS spectra span the entire
optical domain (from 370 to 1,000~nm) at a resolving power of about
65,000.  In total, 32 circular-polarization sequences were collected,
most of them consisting of 4 individual subexposures taken in
different polarimeter configurations (in two cases, the sequence was
interrupted after the second exposure due to a technical problem with
the instrument control).  The extraction procedure is described in
\citet[][see Donati et al., 2006, in prep., for further
details]{Donati97} and was carried out with Libre~ESpRIT, a fully
automatic reduction package/pipeline installed at CFHT for optimal
extraction of ESPaDOnS spectra.  The peak signal-to-noise ratios per
2.6~\kms\ velocity bin range from 490 to 1710, depending mostly on
weather conditions and exposure time (see
Table~\ref{tab:log})\footnote{Note that the instrument suffered a
1.3-mag loss in throughput compared to the optimal performance from
early March to late June 2005, due to severe damage to the external
jacket of optical fibres linking the polarimeter with the spectrograph
(now fixed).}.

Three additional circular-polarization spectra were obtained at the
3.9-m Anglo-Australian Telescope (AAT) in late Sep.~2004, using a
visitor-instrument Cassegrain polarimeter, fibre linked to the UCL
echelle spectrograph (UCLES), in a setup very similar to that
described by \citet{Donati03}.  The AAT spectra, processed with the
same reduction package as used for the CFHT data, cover 430--670~nm at
a resolving power of 65,000.  The peak signal-to-noise ratios per
2.6~\kms\ velocity bin range from 670 to 890, depending on the 
weather (see Table~\ref{tab:log}).

Both instruments use Fresnel rhombs (rather than crystalline plates)
as retarders, with the result that spectropolarimetric ripples
\citep[e.g.,][]{Aitken01, Semel03} are decreased down to a level below
detectability.  An example of this improvement compared to previous
otherwise similar instruments (such as the MuSiCoS spectropolarimeter;
\citealt{Donati99}) is illustrated by \citet{Wade05} and
\citet{Wade06} in the particular case of hot-star observations.

In order to gain a multiplexing advantage by combining results from
different lines in the spectra, Least-Squares Deconvolution (LSD;
\citealt{Donati97}) was applied to all observations.  The line list
required for LSD was computed from an Atlas9 LTE model atmosphere
\citep{Kurucz93} at $\teff=30,000$~K, $\logg=4.5$, roughly matching
$\tau$~Sco's parameters.  We utilized only moderately strong lines
(those with synthetic profiles having line-to-continuum core
depressions larger than 10\% prior to all non-thermal broadening
mechanisms, but omitting the very strongest, broadest features, such
as Balmer and He lines, whose Zeeman signature is strongly smeared out 
compared to those of narrow lines) -- some 500 spectral features altogether, 
with about half corresponding to oxygen lines and most others coming from
N, Si, C and Fe.  The average noise levels in the resulting LSD
signatures range 0.6--2.1$\times10^{-4}$ for ESPaDOnS spectra and
1.0--1.4$\times10^{-4}$ for the AAT data (per 1.8-\kms\ velocity bin, relative
to the unpolarized continuum level $I_{\rm c}$; see
Table~\ref{tab:log}).  Significant Zeeman signatures are clearly
detected in all the spectra, with a full amplitudes of about 0.2\%,
demonstrating that a magnetic field is securely detected for
$\tau$~Sco; an illustrative example is shown in Fig.~\ref{fig:lsd}.

\begin{figure}
\center{\includegraphics[scale=0.35,angle=-90]{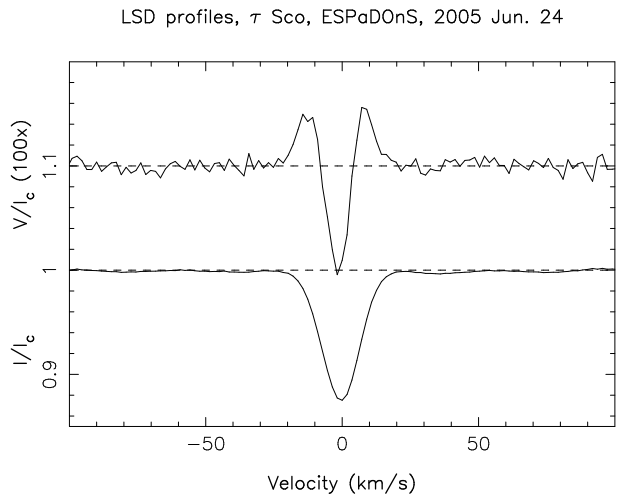}} 
\caption[]{LSD unpolarized and circularly polarized profiles of
$\tau$~Sco (bottom, top curves respectively), as observed on 2005
June~24.  The mean polarization profile is expanded by a factor of 100
and shifted upwards by 1.10 for display purposes.  }
\label{fig:lsd}
\end{figure}

The Zeeman signatures detected are variable with time\footnote{The LSD
$I$ profiles are also variable in strength. The level of this
variability is, however, small (about 0.3\% rms of the continuum
level, equivalent to $\sim$3\%\ of the central line depth), affecting mainly
the line width.  We return to this point in Sec.~\ref{sec:params}.  }  
and correspond to projected longitudinal fields
(computed from the first-order moment of the Stokes $V$ LSD profile;
\citealt{Donati97}) ranging from $-50$ to +90~G (see
Table~\ref{tab:log}).  Our results are compatible with the earlier
findings of \citet{Landstreet82} that longitudinal fields of normal
upper-main-sequence stars in general, and of $\tau$~Sco in particular
(for which \citealt{Landstreet82} report two estimates), are less than
100~G on average.

\subsection{UV spectroscopy}

We augmented the new spectro{\-}polarimetric data set with archival UV
spectra obtained with the International Ultraviolet Explorer satellite
\citep[IUE;][]{Boggess78a}.  The IUE archive contains 107
high-resolution spectra of $\tau$~Sco, covering the wavelength range
115--195~nm, sampled at 0.01-nm intervals with a resolution of about
$10^4$ and $\sn \simeq 20$.  We examined spectra obtained from the
MAST IUE archive at the Space Telescope Science Institute, reduced
using {\sc newsips} pipeline software \citep[][the {\sc `mxhi'}
product]{Nichols96}.  IUE spectra were each acquired through one of
two focal-plane apertures, designated `small' and `large'; we found
indications of data-processing errors in a number of small-aperture
spectra, which also generally have poorer signal and \sn\ than the
large-aperture data (typically, the small aperture transmitted only
$\sim$half the incident radiation).  We therefore present results
based on only the 72 large-aperture spectra in the archive;
incorporating the small-aperture data introduces no material changes
to the conclusions.

\section{The rotation period of $\tau$~Sco}
\label{sec:rot}

\subsection{Zeeman spectroscopy}
\label{sec:rotZmn}

Time variability of the projected longitudinal field is a potentially
powerful tool for estimating stellar rotation periods, which often
cannot be measured by other means.  Specifically, temporal
fluctuations of the line-of-sight component of the field in hot stars
may be attributed to a magnetic topology that is not axisymmetric
about the rotation axis, thus showing different configurations to the
observer as the star rotates.  Searching for timescales on which the
longitudinal field repeats identically from one cycle to the
next has been very successful in estimating rotation periods of Ap
and Bp stars \citep[e.g.,][]{Borra80}.

Of course, the longitudinal-field values contain far less information
about the field topology than the Zeeman signatures themselves.  The
Stokes $V$ LSD profile shown in Fig.~\ref{fig:lsd} provides a clear
demonstration of this point; while the Zeeman signature strongly
indicates a field detection (at a level of about 27$\sigma$), the
corresponding longitudinal field is $1.3\pm3.1$~G and is therefore
inconclusive by itself.  Nonetheless, longitudinal-field values are
convenient summary statistics in several respects, and in particular
can provide useful rough estimates of the average magnetic flux over
the stellar surface.

As an initial step in modelling the longitudinal-field variations of
$\tau$~Sco, we compared a double sine-wave fit to the data, with the
two periods held in the ratio 2:1 (motivated by what we expect from a
simple, rotationally modulated, linear combination of dipole plus 
quadrupole fields).  For a range of assumed
values for the longer (i.e., rotational) period, the amplitudes and
phases of both waves were optimized to obtain the best fit to the
longitudinal-field measurements, using a standard least-squares
minimization process, with the reduced-$\chisq$ value, \chinu,
evaluated as the statistic of merit for the fit quality.
Fig.~\ref{fig:beff} (upper panel) shows the results of this exercise,
and indicates a strong minimum in \chinu\ at $P_{\rm rot} =
41.08\pm0.07$~d (1$\sigma$ uncertainty).

However, the double sine-wave model evidently provides no more than a
rough fit to the data, with a minimum \chinu\ as large as $\sim$9.
This, in turn, indicates that the field must be significantly more complex
than a tilted dipole plus quadrupole combination, and must contain higher-order, 
multipole components.\footnote{The rotational modulation of the
longitudinal field is very reminiscent of what is observed on the
helium-strong star HD~37776 \citep{Thompson85}, apart from the fact
that the field of HD~37776 is stronger than that of $\tau$~Sco by more
than an order of magnitude.}  Clearly, more-detailed modelling of the
Zeeman signatures is necessary in order to obtain an adequate
description of the magnetic topology of $\tau$~Sco; we describe such
modelling in Sec.~\ref{sec:mod}.

The mis-match between the simple model and the data means that the
41-d signal is no more than a provisional estimate of
the rotation period.  More accurate and precise values may be obtained
both from modelling the Zeeman signatures directly
(Sec.~\ref{sec:mod}), and from a time-series analysis of the IUE data
(Sec.~\ref{sec:rotIUE}).

\begin{figure}
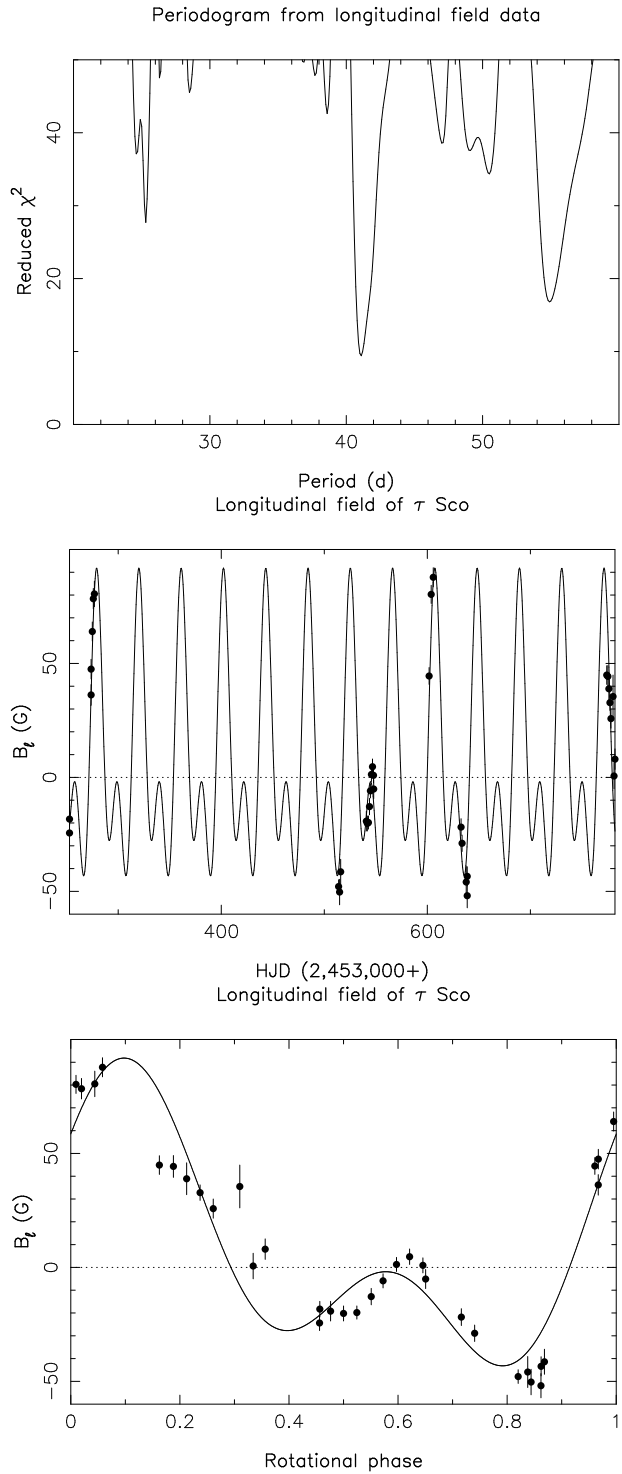

\center{\includegraphics[scale=0.35,angle=-90]{fig/tausco_beff3.ps} 
        \includegraphics[scale=0.35,angle=-90]{fig/tausco_beff1.ps} 
        \includegraphics[scale=0.35,angle=-90]{fig/tausco_beff2.ps}} 
\caption[]{Periodogram resulting from a double sine-wave fit to the
longitudinal-field data of $\tau$~Sco.  {\em Top panel:} \chinu\ as a
function of the period of the main sine wave.  A clear minimum is
obtained for a period of about 41~d.  {\em Middle panel:} Temporal
fluctuations of the longitudinal field of $\tau$~Sco (full dots, with
1$\sigma$ error bars) along with the model fit (full line) for the
adopted period of 41.033~d, as a function of heliocentric Julian date
(HJD).  {\em Bottom panel:} As for the middle panel, but as a function
of rotation phase, computed using the ephemeris of
eqn.~\ref{eq:ephem}.}
\label{fig:beff}
\end{figure}

\begin{figure*}
\center{\includegraphics[scale=0.7 ,angle=-90]{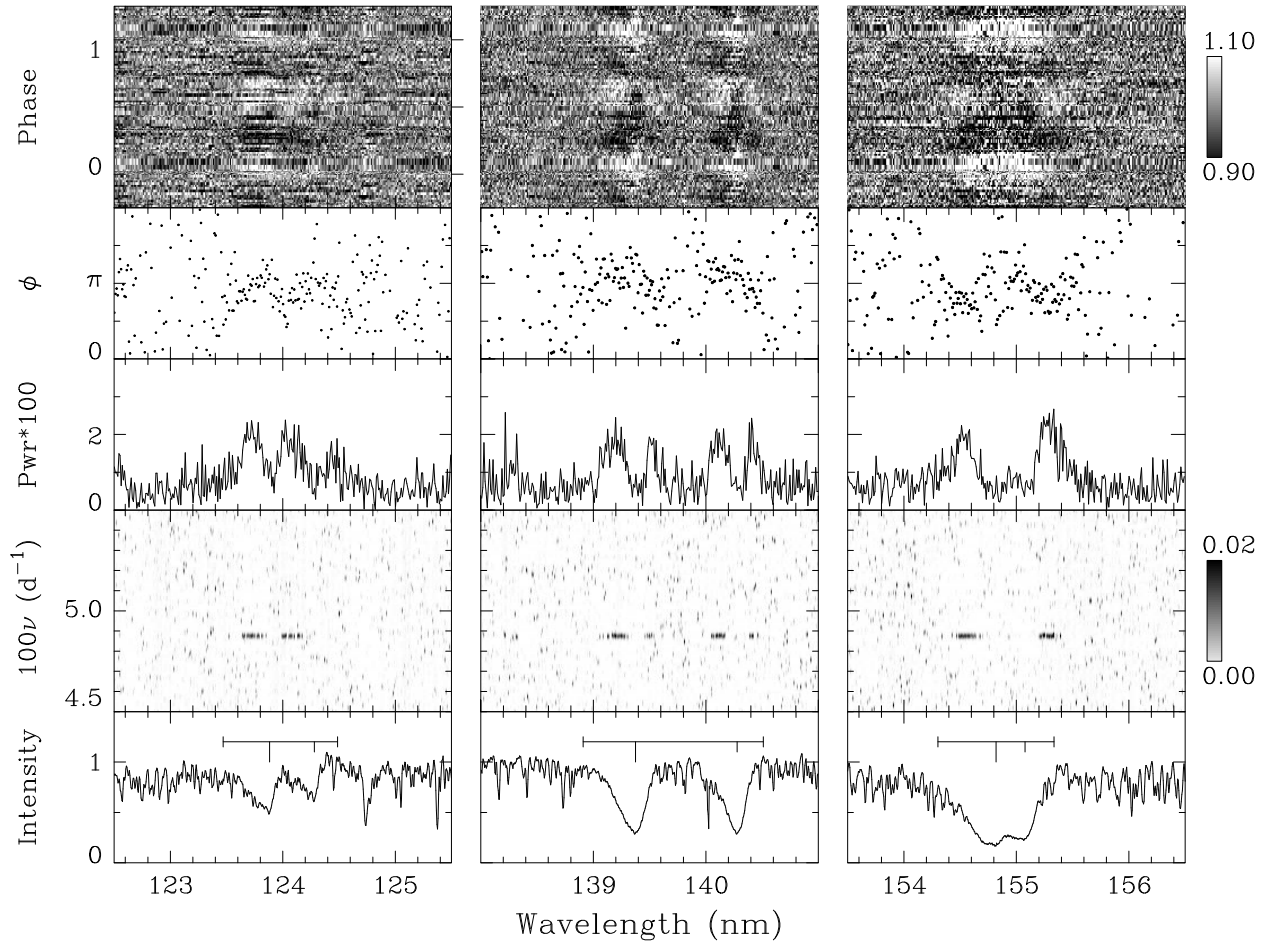}}
\caption[]{Time-series analysis of IUE spectra of $\tau$~Sco.  The
bottom panel shows the mean spectrum for (left to right) the \nv,
\siiv, and \civ\ resonance doublets; in each case, vertical tickmarks
indicate the rest wavelengths, while the horizontal bar extends from
1000~\kms\ bluewards of the short-wavelength component to +500~\kms\
redwards of the long-wavelength component.   The next panel up shows part
of the 2D {\sc clean}ed periodogram, showing evidence for a periodic
signal at $\nu = 0.049$~d$^{-1}$ in the wind-formed lines
(see Fig.~\ref{fig:UVfig2}), while the next two
panels show the fourier power and phase, respectively, at that
frequency.  The top panel shows the residuals about the mean of the rectified
spectra phased according to the ephemeris of eqn.~\ref{eq:ephem}. }
\label{fig:UVfig1}
\end{figure*}

\subsection{UV line-profile variability}
\label{sec:rotIUE}

Direct measurement of magnetic fields in hot stars
\citep[e.g.;][]{Donati01, Donati02, Neiner03, Donati06} has often been
presaged by the detection of strictly periodic UV line-profile
variability, with the inference of a magnetic field by analogy with He
peculiar stars \citep[e.g.;][]{Stahl96, Neiner03, Walborn04}.  A
search for variability in the IUE spectra of $\tau$~Sco (known to exhibit 
abnormally strong UV P-Cygni lines; \citealt{Walborn84}) 
%
%
%
is therefore a natural follow-up to 
our detection of rotationally modulated Zeeman signatures in the 
spectro{\-}polarimetric data.  

We are fortunate that
$\tau$~Sco was adopted as a photo{\-}metric calibration star by the
IUE Project, and that as a consequence the temporal sampling of the UV
spectra is rather well suited to investigating the rotational
timescale identified in the Zeeman spectroscopy: the large-aperture
spectra span 16.6~yr (1979~Feb -- 1995~Sept, median date 1988~Aug) at
a median sampling rate of 50.7~d (range 0.7~hr -- 417~d).

\begin{figure}
\center{\includegraphics[scale=0.35 ,angle=-90]{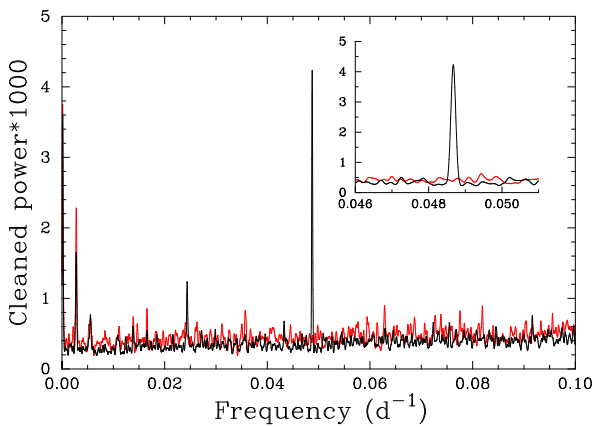}}
\caption[]{Mean power spectrum for regions extending
1000~\kms\ bluewards of the short-wavelength component to +500~\kms\
redwards of the long-wavelength component for the \nv~124~nm,
\siiv~140~nm, and \civ~155~nm resonance doublets (black line), and for
equivalent regions 2~nm longwards (red line).}
\label{fig:UVfig2}
\end{figure}

We analyzed the rectified IUE spectra using the implementation of the
{\sc clean} algorithm described by \citet{Roberts87}; some results of
this analysis are presented in Fig.~\ref{fig:UVfig1}.  There is a
strong signal at $\nu=0.049$~d$^{-1}$ ($P=20.5$~d) in the main
features formed in the stellar wind (the \nv~124~nm, \siiv~140~nm, and
\civ~155~nm resonance lines, but also \siiii~120.6~nm).  To examine
this in greater detail, we constructed the mean power spectrum for
wavelengths from 1000~\kms\ bluewards of the short-wavelength
component to 500~\kms\ redwards of the long-wavelength component for
each of the three resonance doublets, together with a comparison
spectrum from equivalent regions 2~nm longwards of each feature;
results are shown in Fig.~\ref{fig:UVfig2}.

In addition to the main signal, there is a weak secondary signal at
$\nu=0.024$~d$^{-1}$ (Fig.~\ref{fig:UVfig2}), corresponding to a period 
of about 41~d.\footnote{There are also
signals, in both the main and reference datasets, at $P=0.5$~yr and
1~yr.  These are presumed to be non-astrophysical (and are most
probably the result of varying background levels).}  Since a period of
20.5~d is completely excluded by the Zeeman measurements
(Sec.~\ref{sec:rotZmn}), we interpret our findings as indicating a
slightly non-sinusoidal double wave with $P_{\rm rot}=41.03$~d.  To
refine this period we average results from the 75 0.1-\AA\ wavelength
bins in the $-1000/+500$~\kms\ test ranges for which the peak of the
power spectrum is safely identifiable with the $\nu = 0.049$~d$^{-1}$
signal;  half the mean peak frequency for those samples corresponds to
$P_{\rm rot}=41.033\pm0.002$~d (s.e.).

\begin{figure}
\center{\includegraphics[scale=0.35 ,angle=-90]{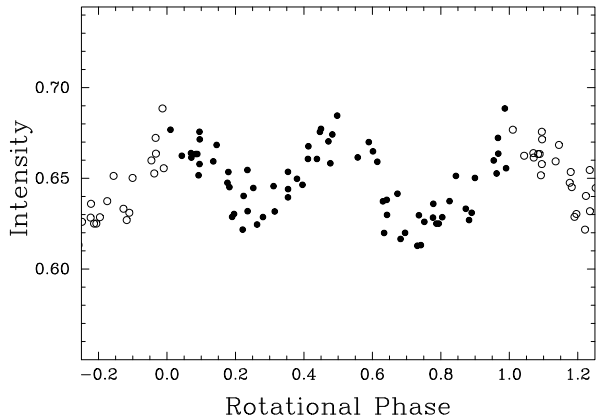}}
\caption[]{Mean residual intensity for regions from 1000~\kms\ bluewards of the 
short-wavelength component to +500~\kms\ redwards of the long-wavelength component 
for each of the \nv~124~nm, \siiv~140~nm, and \civ~155~nm resonance
doublets, as a function of rotational phase.  Phases are computed according to the 
ephemeris of eqn.~\ref{eq:ephem}. 
}
\label{fig:UVfig3}
\end{figure}

As an alternative characterization of the data, fitting a gaussian to
the peak of the mean power spectrum yields $P{\rm rot}=41.034$~d, and
a full-width at half-maximum (FWHM) of 0.36~d.  To obtain a second, reasonably conservative,
estimate of the uncertainty on the period, we may assume that the
signal does not go out of phase by more than 0.125$P_{\rm rot}$ (i.e.,
0.25 of the 20.5-d half-wave period) in each of the contributing
wavelengths, giving $P = 41.0340 \pm 0.0040$~d (s.e.).

We emphasize that the data are fully consistent with a strictly
periodic signal; the IUE data, spanning almost 17~yr, show no evidence
for any non-periodic component insofar as the width of the periodogram
peak is entirely accounted for by the finite number of cycles in the
dataset ($N=148$).  Moreover, the same period is recovered, to high
accuracy, in completely independent datasets separated by a
quarter-century (viz., the IUE and Zeeman spectroscopy; more-detailed
modelling of the Zeeman data, reported in Sec.~\ref{sec:mod}, yields
$P_{\rm rot} = 41.02\pm0.03$~d).  Furthermore, the precision of the
adopted rotation period is high enough to ensure that the relative
phasing between the archival IUE spectra and our new
spectro{\-}polarimetric data is better than 1\%.

On the basis of these results we therefore adopt
\begin{equation}
T_0 = \mbox{HJD~}2,453,193.0  + 41.033({\pm}0.002)E
\label{eq:ephem}
\end{equation}
as our rotational ephemeris, where phase zero is arbitrarily chosen
for convenience as a date just prior to the acquisition of the first
Zeeman data.  The rotational modulation of the mean residual intensity
over the \nv, \siiv\ and \civ\ UV resonance doublets with this
ephemeris is shown in Fig.~\ref{fig:UVfig3}; significant absorption
episodes are observed to occur in all 3 spectral features around phase
0.3 and 0.8.

\section{Rotational properties}
\label{sec:params}

From our analysis, $\tau$~Sco is the second-slowest rotator so far known 
among high-mass stars (cp.\ HD~191612, whose rotation period is
suggested to be 538~d; \citealt{Donati06}).  In order to be able to
refine our modelling of its magnetic topology, we first summarize what
we know about its physical parameters, and examine how these tie in
with the rotation period we derive.

From detailed spectroscopic studies by \citet{Kilian92},
\cite{Mokiem05} and \cite{SimonDiaz05}, we know that $\tau$~Sco is a
young star, with an age of a few
Myr\footnote{Although $\tau$~Sco is quoted to be younger than 1~Myr by
\citet{Kilian92}, its membership in the Sco~OB2 association
\citep{deZeeuw99} suggests that $\tau$~Sco, as the most massive star
in the group, is roughly as old as the group itself -- i.e., nearly
5~Myr.}, a mass of  $\sim$15~\msun, a temperature of $31,500\pm500$~K, and a radius of
$5.2\pm0.5$~\rsun.\footnote{Significantly larger radii are cited in
the literature, e.g.\ by \citealt{Howk00}, but they are not consistent
with the well-established Hipparcos parallax.}  The radius and
rotation period imply an equatorial rotation velocity, \ve, of only
$6.4\pm0.6$~\kms, and thus a line-of-sight projected rotation
velocity, \vesini, which is even less (where $i$ is the inclination of the
rotation axis to the line of sight).  This \vesini\ limit is
significantly lower than the estimate given by \citet{Kilian92} from
line-profile modelling (19~\kms), but is similar the value (of 5~\kms) 
reported by both \citet[][]{Smith78} and \citet[][]{Mokiem05}, and is 
consistent with the upper limit given by \citet[][13~\kms]{SimonDiaz05}.

We performed our own simple modelling of the LSD Stokes $I$ profile,
to get an estimate of how much (and what type of) broadening is
required in addition to a pure rotational broadening in order to
obtain a reasonable fit to the data.  We find that gaussian broadening
with an average FWHM of 13.5~\kms\ gives a nice match to
the mean LSD profile (after allowance for a gaussian instrumental 
broadening, with a FWHM of 5~\kms).   
%
%
This corresponds to a turbulent
velocity of about 8~\kms, a value similar to estimates given  by
\citet{Mokiem05} and \citet[][10.8 and 8.7~\kms,
respectively]{SimonDiaz05}.  This broadening is larger than the
thermal broadening expected for most lines used in the LSD process
(about 5~\kms\ in average, ranging from 3 to 6~\kms, depending on the
atom), and the similar widths found for lines from different atoms
confirm that the main origin is non-thermal.  

\begin{figure}
\center{\includegraphics[scale=0.5]{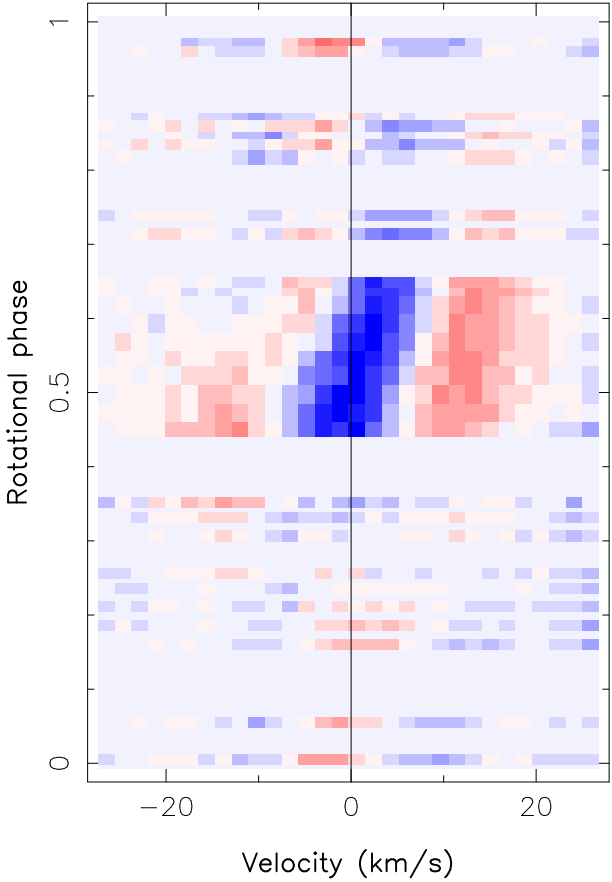}} 
\caption[]{Dynamic LSD Stokes $I$ spectrum of $\tau$~Sco, demonstrating that 
the average unpolarised line is also modulated with the detected rotation 
period of 41~d (see Sec.~\ref{sec:rot}).  
A simple synthetic profile (featuring gaussian turbulence broadening with 
a FWHM of 13.5~\kms\ and $\vesini=6$~\kms, see text) was removed from all profiles 
to emphasize variability.  A clear signal, centred at phase 0.5, crosses the line 
profile and indicates that the average photospheric line of $\tau$~Sco gets slightly 
narrower (by about 10\%) around phase 0.5.  Blue and red respectively correspond to 
line absorption/emission features with an amplitude of $\pm0.4\%$ of the continuum 
intensity (i.e.\ $\pm3\%$ of the central line depth). }
\label{fig:dyni}
\end{figure}

As already mentioned in Sec.~\ref{sec:obs}, we observe that the Stokes
$I$ LSD profiles of $\tau$~Sco are slightly variable with time, and
that this temporal variation can also be attributed as modulation with
the 41~d rotation period derived at Sec.~\ref{sec:rot} (see
Fig.~\ref{fig:dyni}).  This modulation consists mainly of a small
$\sim$10\%), phase-locked variation of the line width, with the
average photospheric profile being slightly narrower around phase 0.5.
We suspect that this indicates that the atmospheric turbulence evinced
by the line broadening is induced at the base of the wind, and is not
constant over the surface of $\tau$~Sco, reflecting, perhaps, the
spatial distribution of field strengths and orientations; another
possibility is that the density at the base of the wind is not
constant over the surface of the star (again, as a potential result of
the magnetic field), causing the observer to see to different
atmospheric depths (and thus different turbulent broadenings) at
different rotational phases.  We also find that a slightly better fit
to the Stokes $V$ signatures is obtained when assuming a local line
profile width about 10\% narrower than that derived from LSD Stokes
$I$ profiles; we speculate that Zeeman signatures are formed at
atmospheric depths slightly different from the unpolarised spectral
lines, i.e.\ in a layer where turbulent broadening is weaker.

From the observed variability of the longitudinal field, which in
particular shows a steep gradient as well as a sign switch in as
little as 10\% of the rotational cycle (between phase 0.85 and 0.95;
see Fig.~\ref{fig:beff}, bottom panel), we conclude that it is
unlikely that $\tau$~Sco is seen near-pole-on; modelling of the Zeeman
signatures (Sec.~\ref{sec:mod}) reveals that significantly better
fits to the data are obtained with $\vesini=6$~\kms\ than with values
$\le$5~\kms.  This implies that $i$ is large, probably in the range
60\degr--90\degr.  Since there is only a low probability that the star
is seen exactly equator on, we assume $i=70\degr$ (and
$\vesini=6$~\kms) in the following analysis.  Very similar results are 
obtained with values of $i$ ranging from 60\degr\ to 90\degr.  

Finally, we note that the radial velocity of $\tau$~Sco has remained
remarkably stable during the whole length of our run, at
$-0.6\pm0.1$~\kms.  This confirms earlier conclusions by
\citet{Stickland95} that there is no reason to suppose that $\tau$~Sco
a member of a binary system.
%
%

\section{Modelling the surface magnetic topology of $\tau$~Sco}
\label{sec:mod}

\subsection{Methodology}

To reconstruct the surface magnetic topology of $\tau$~Sco from the
set of observed Zeeman signatures, we use our magnetic-imaging code
\citep{Brown91, Donati97b} in its most recent implementation
\citep{Donati01b}.  While still based on maximum-entropy image
reconstruction, this latest version reconstructs the field topology as
a spherical-harmonic decomposition, rather than as a series of
independent magnetic-image pixels as before.  One obvious advantage of
this method is that we can impose {\em a priori} constraints on the
field topology -- e.g., that the field is purely potential, or purely
toroidal, or a combination of both.  Another important advantage of
this formalism is that both simple and complex magnetic topologies can
easily be reconstructed \citep{Donati01b}, whereas the original method
failed at reconstructing simple magnetic geometries \citep[such as
dipoles;][]{Brown91}.

To effect this approach, we describe the field as the sum of a
potential and a toroidal component, each expressed as a
spherical-harmonic expansion.  In a formalism similar to that of
\citet{Jardine99}, the field components can be written as:
\begin{eqnarray}
B_{r}(\theta,\phi)      & = & - \sum_{\ell,m} \alpha_{\ell,m} Y_{\ell,m}(\theta,\phi)  \\ 
B_{\theta}(\theta,\phi) & = & - \sum_{\ell,m} \left( \beta_{\ell,m} Z_{\ell,m}(\theta,\phi) 
                                    + \gamma_{\ell,m} X_{\ell,m}(\theta,\phi) \right)  \\ 
B_{\phi}(\theta,\phi)   & = & - \sum_{\ell,m} \left( \beta_{\ell,m} X_{\ell,m}(\theta,\phi) 
                                    - \gamma_{\ell,m} Z_{\ell,m}(\theta,\phi) \right),
\end{eqnarray}
where 
\begin{eqnarray}
Y_{\ell,m}(\theta,\phi) & = & c_{\ell,m} P_{\ell,m}(\theta)\ e^{i m \phi}  \\ 
Z_{\ell,m}(\theta,\phi) & = & \frac{c_{\ell,m}}{\ell+1} \frac{\partial P_{\ell,m}(\theta)}{\partial \theta}\ e^{i m \phi}  \\ 
X_{\ell,m}(\theta,\phi) & = & \frac{c_{\ell,m}}{\ell+1} \frac{P_{\ell,m}(\theta)}{\sin \theta}\ i m\ e^{i m \phi}  \\ 
c_{\ell,m} & = & \sqrt{\frac{2 \ell + 1}{4 \pi}\frac{(\ell-m)!}{(\ell+m)!}}, 
\end{eqnarray}
with $\ell$ and $m$ denoting the order and degree of the
spherical-harmonic mode $Y_{\ell,m}(\theta,\phi)$ ($\theta$ and $\phi$
being the colatitude and longitude at the surface of the star), and
$P_{\ell,m}(\theta)$ the associated Legendre polynomial.  For a given
set of the complex coefficients $\alpha_{\ell,m}$, $\beta_{\ell,m}$
and $\gamma_{\ell,m}$ (where $\alpha_{\ell,m}$ characterises the 
radial field component, $\beta_{\ell,m}$ the azimuthal and meridional 
components of the potential field term, and $\gamma_{\ell,m}$ the azimuthal 
and meridional components of the toroidal field term), one can produce the 
associated magnetic image
at the surface of the star, and thus derive the corresponding Stokes
$V$ dataset.  We carry out the inverse problem, aimed at
reconstructing a set of complex coefficients from an automated,
iterative fit to the observed circular-polarization LSD profiles.
Principles of maximum-entropy image reconstruction are applied to the
set of complex coefficients, rather than on the image pixels.  This is
similar to what is presented by \citet{Hussain01}, except that we
generalize the problem to fields that are non-potential and feature a
significant toroidal component.  Fitting a pure potential field to the
data is equivalent to fitting $\alpha_{\ell,m}$ and $\beta_{\ell,m}$
alone (setting all $\gamma_{\ell,m}$ to zero); using all three sets of
coefficients in the fitting procedure produces a more general magnetic
topology, with a non-zero toroidal field.  Trying both approaches is a
straightforward way of investigating whether or not the magnetic field at the
surface of $\tau$~Sco is potential in nature.

\begin{figure*}
\center{\includegraphics[scale=0.9,angle=-90]{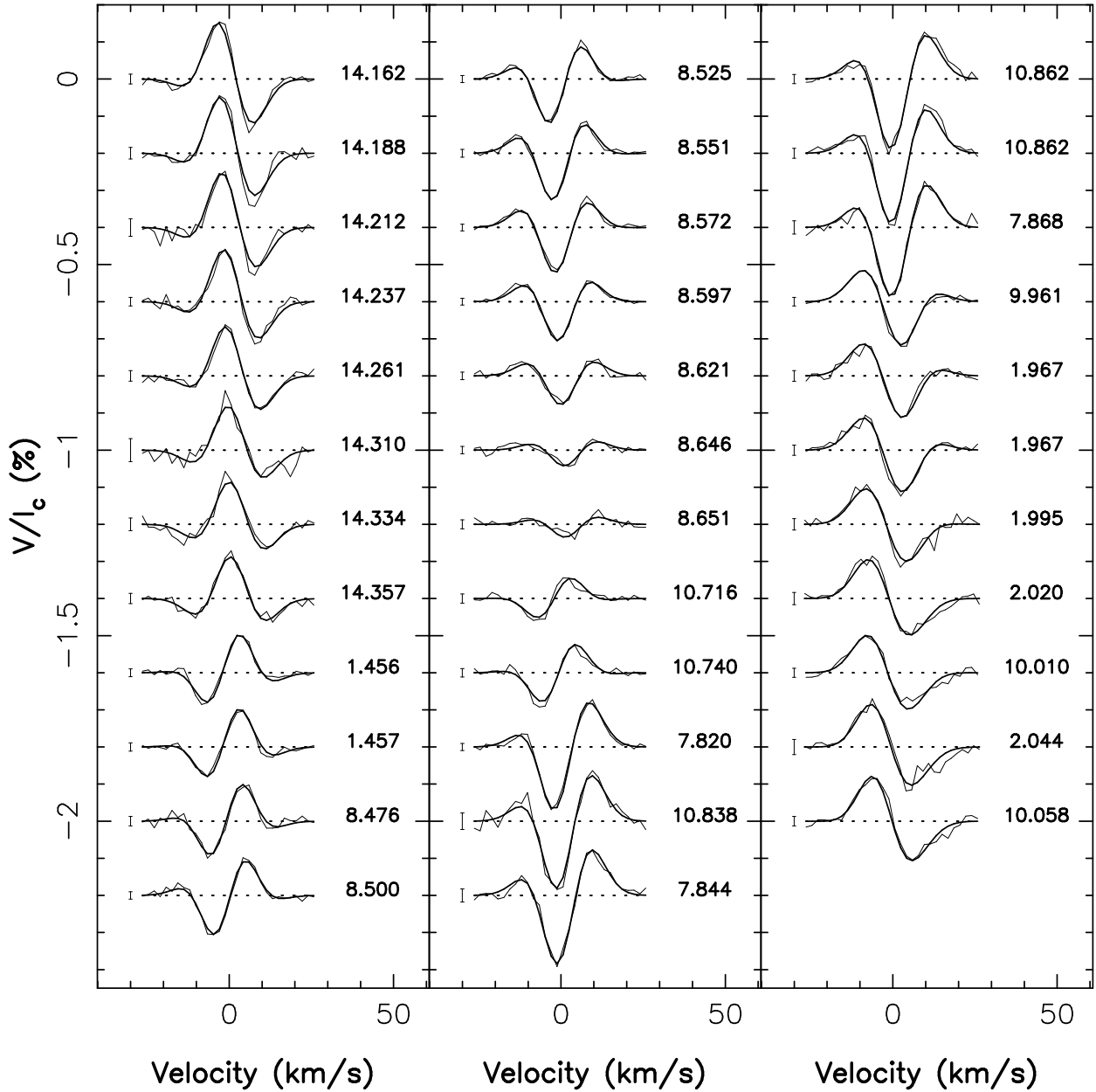}} 
\caption[]{Maximum-entropy fit (thick lines) to the observed Zeeman
signatures of $\tau$~Sco (thin lines).  The rotational phase and cycle
of each observation is written next to each profile.  A 3$\sigma$ error 
bar is also plotted left to each profile.  }
\label{fig:fit}
\end{figure*}

For a spectral resolution of 5~\kms, a microturbulent velocity of
7.5~\kms, and a projected equatorial velocity \vesini\ of 6~\kms, the
number of resolved equatorial elements around the star is about 5,
implying that we need about 10 equatorial elements to reproduce the
observations with adequate surface sampling.  Truncating the
spherical-harmonic expansion of the magnetic-field components to terms
with $\ell\leq 10$ is therefore sufficient in the case of $\tau$~Sco,
and should introduce negligible degradation in the spatial resolution
of our reconstructed images.  In practice, we used $\ell\leq 12$ and
found that, as expected, no improvement in the quality of the fit to
the data was obtained when adding higher-order terms.  This
corresponds to mapping a total of 90 modes at the surface of the star,
implying a total of 360 image parameters in the case of a potential
field, and 540 in the case of a potential- plus toroidal-field
configuration.

We first carried out a series of magnetic reconstructions for a wide
range of values of the rotation period (without constraining the field
to a specific type of configuration).  The minimum \chinu\ is obtained
at $P_{\rm rot} = 41.02\pm0.03$~d (1$\sigma$ uncertainty).  This
estimate of $P_{\rm rot}$ is a refinement of that derived in
Sec.~\ref{sec:rotZmn}, because of the more complete physical model; it
is fully compatible with the adopted, more precise period derived
independently from the IUE data (eqn.~\ref{eq:ephem};
Sec.~\ref{sec:rotIUE}).

The best-fit model of the Zeeman signatures is shown in
Fig.~\ref{fig:fit}, from which it is obvious that the greater part of
the observed profile information is satisfactorily reproduced.
Nonetheless, the minimum \chinu\ is as large as 1.5, indicating that
discrepancies between the model and observations still exist.  The
origin of these (small) discrepancies is not yet clear, but may result
from the simple isotropic local line-profile model we use to compute
the synthetic Stokes $V$ profiles (Sec.~\ref{sec:params}).
We note that the fit to the data is much worse if we force the 
magnetic topology to be very simple, e.g., similar to that found in 
most magnetic chemically peculiar stars to date.  For instance, when 
truncating the spherical harmonics expansion to $\ell\leq 1$ (equivalent 
to fitting the data with a tilted magnetic dipole model), the minimum 
achievable \chinu\ is 15;  with $\ell\leq 2$ (roughly equivalent to 
adding up a magnetic quadrupole component to the model), the fit 
quality is still very rough ($\chinu\simeq8$).  The detected Stokes $V$ 
profiles (and in particular the observed rotational modulation) 
definitely indicate that the magnetic-field topology of $\tau$~Sco is 
much more complex than usual (by massive star standards).  

\begin{figure}
\center{\includegraphics[scale=0.85]{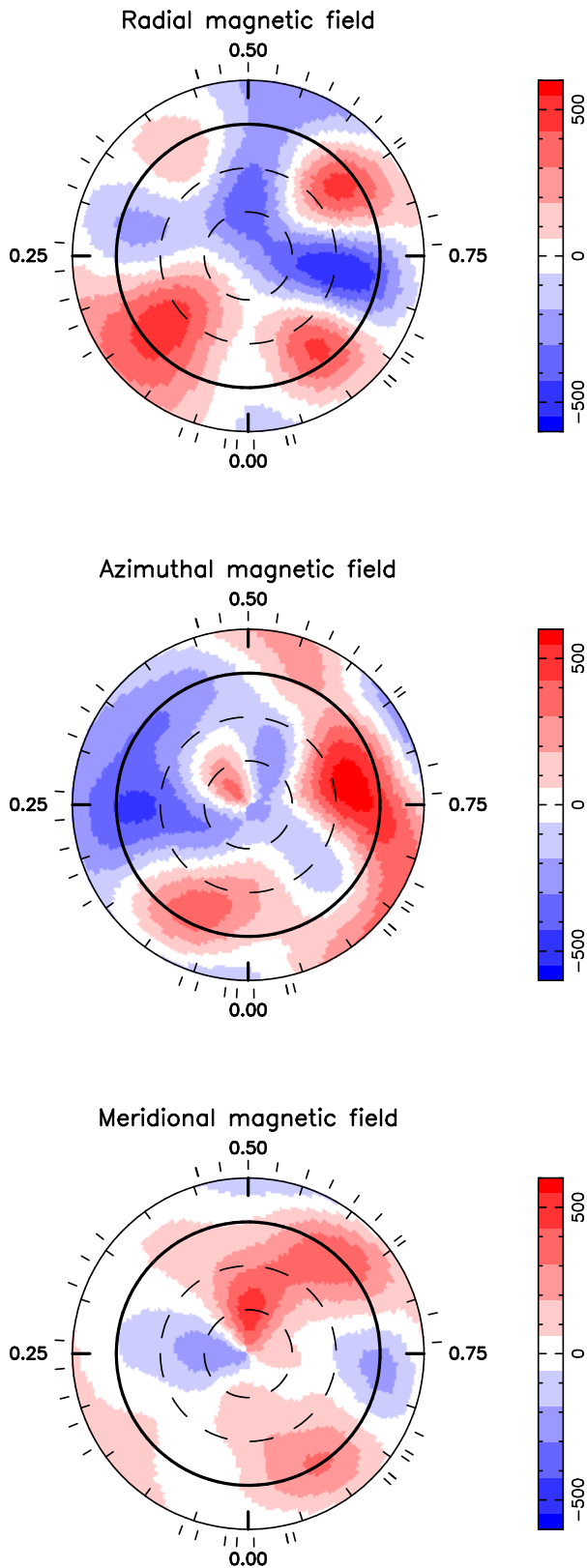}} 
\caption[]{Maximum-entropy reconstructions of the magnetic topology of
$\tau$~Sco, assuming that the global field can be expressed as the sum of a 
potential field and a toroidal field.  The three components of the field 
are displayed from top to bottom (flux values labelled in G).  
The top image (radial field component) is described through the set of 
complex coefficients $\alpha_{\ell,m}$ (see Sec.~\ref{sec:mod}).   
The star is shown in flattened polar projection down to latitudes
of $-30\degr$, with the equator depicted as a bold circle and
parallels as dashed circles.  Radial ticks around each plot indicate
phases of observations.  }
\label{fig:map}
\end{figure}

\begin{figure*}
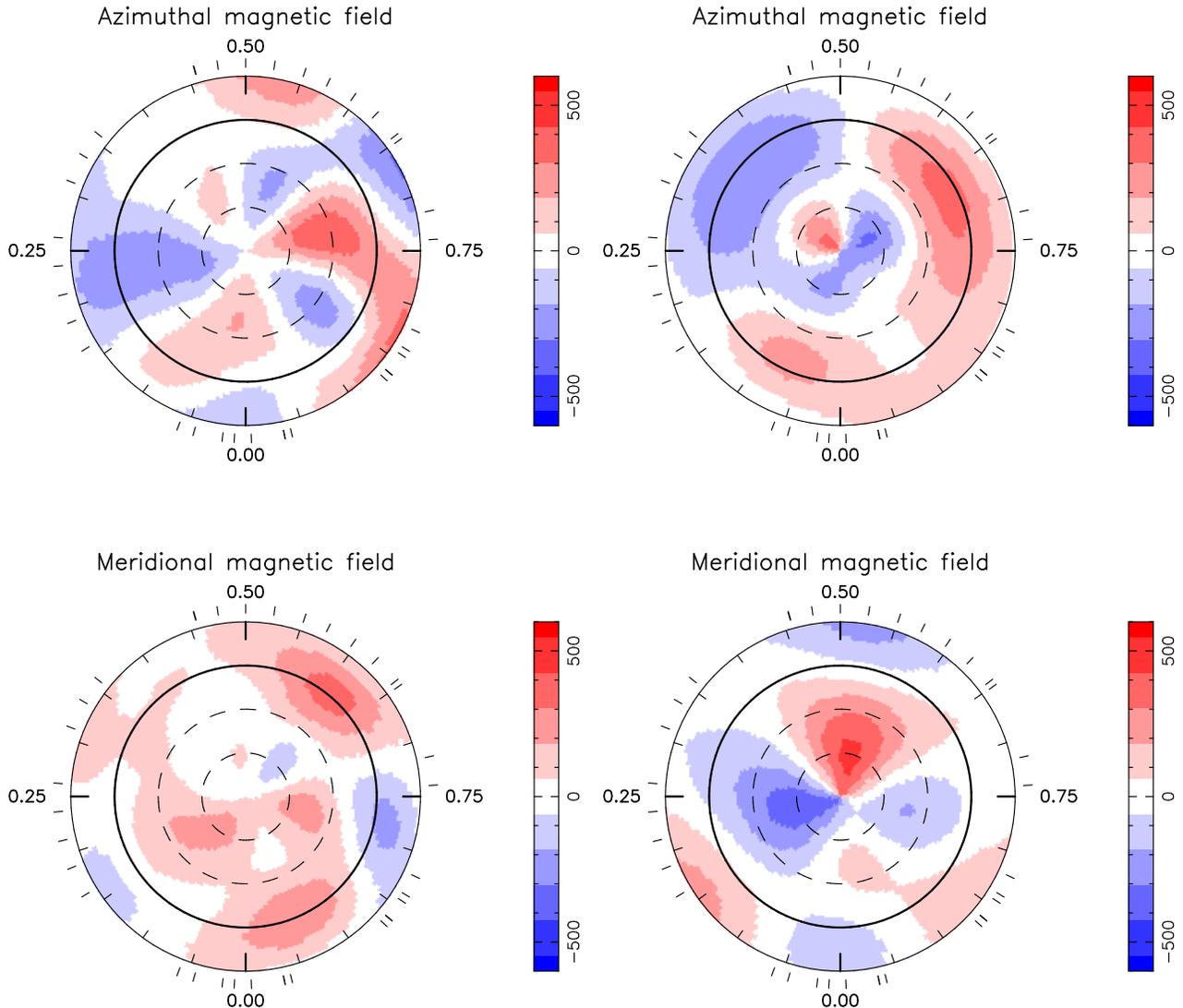

\center{\mbox{\includegraphics[scale=0.59]{fig/tausco_map3.ps}\hspace{2mm}
              \includegraphics[scale=0.59]{fig/tausco_map4.ps}}} 
\caption[]{Azimuthal and meridional components of the reconstructed 
potential (left column) and toroidal (right column) field structures. 
Adding both together yields the azimuthal and meridional field components 
shown in Fig.~\ref{fig:map}.  The image on the left hand side is described 
through the set of complex coefficients $\beta_{\ell,m}$ while that on the 
right hand side is obtained through the coefficients $\gamma_{\ell,m}$ (see 
Sec.~\ref{sec:mod}).  }
\label{fig:map2}
\end{figure*}

\subsection{Results}

The magnetic topology we reconstruct by assuming that the field
includes both a potential field and a toroidal field component is
shown in Fig.~\ref{fig:map}.  The corresponding \chinu\ is 1.5.  If we
instead assume that the field is purely potential, the optimal fit we
obtain yields $\chinu=1.8$, indicating a slightly poorer fit; while
the inferred topology is broadly similar to that shown in
Fig.~\ref{fig:map}, the contrast and information content is 
significantly higher than in the previous case.  We can thus say that,
at given image information content, the potential- plus toroidal-field
configuration provides a significantly better fit than the pure
potential-field configuration (with respective \chinu\ values of 1.5
and 1.8), indicating that the first option appears significantly more
likely than the latter.

The contributions of potential and toroidal terms to the azimuthal and
meridional field components are shown separately in
Fig.~\ref{fig:map2} (the toroidal term does not contribute to the
radial field component).  We therefore conclude that the surface
magnetic topology of $\tau$~Sco is mainly potential, but apparently
also includes a toroidal field component.  We find that the potential
field component includes about 70\% of the overall reconstructed
magnetic energy, and thus clearly dominates over the toroidal field
component.  In particular, we note that, compared to the potential
component, this toroidal component is much weaker in $\tau$~Sco than
in most cool magnetic stars observed to date, where the toroidal
component largely exceeded the poloidal one \citep[e.g.,][]{Donati03}.
Moreover, we find that this toroidal component is not primarily
axisymmetric (at least about the rotation axis), featuring for
instance two sign switches (at phases 0.52 and 0.18) at equatorial
latitudes (see Fig.~\ref{fig:map2}, top right panel).

The results further emphasize that the reconstructed magnetic field is
far more complex than a simple dipole; for example, the polarity of
the radial field component switches sign six times along the equator
(instead of just twice as expected for a tilted dipole).  A more
quantitative way of considering the complexity of the field is to
examine the relative strengths of the reconstructed spherical-harmonic
coefficients $\alpha_{\ell,m}$, $\beta_{\ell,m}$ and $\gamma_{\ell,m}$
(see Fig.~\ref{fig:modes}).  Even though the modes corresponding to a
tilted dipole are excited ($\ell=1$ and $m=0-1$), a large number of
other modes carry significant power (especially sectoral modes); the
$\ell=m=4$ mode is among the strongest, and its signature can be
readily seen from the magnetic images of Fig.~\ref{fig:map}.
Unsurprisingly, modes with $\ell>6$ contribute little to the
reconstructed image.  The reconstructed toroidal component only shows
up in low-degree (though non-axisymmetric) modes.

Since our $\tau$~Sco spectropolarimetry spans several rotation cycles,
and since several phase ranges were observed more than once (phases
$\sim$0.00, 0.47 and 0.85; see Table~\ref{tab:log}), we can directly
investigate whether the magnetic field of $\tau$~Sco exhibits signs
of variability on a timescale of about a year.  We find that profiles
obtained at very similar phases in different cycles (e.g., at cycles
2.02 and 10.01)  agree to within the noise level.  

We can also estimate how much latitudinal shear the magnetic topology
of $\tau$~Sco experienced between 2004~Sep.\ and 2005~Sep., using the
methods employed for cool stars by \citet{Petit02} and
\citet{Donati03}; we find no evidence for differential rotation at the
surface of $\tau$~Sco, with an upper limit of about 3~\mrpd\ (i.e., at
least 20 times smaller than for the Sun).  This is in agreement with
our finding that the rotational modulation of UV lines, presumably
related to the magnetic topology (Sec.~\ref{sec:disc3a}), is stable over
timescales of decades (Sec.~\ref{sec:rotIUE}).

We emphasize that only moderate surface spatial resolution can be 
obtained for $\tau$~Sco, as a result of its low rotation velocity; 
close bipolar groups, for example, could therefore easily remain undetected 
if present on scales smaller than the resolution element.  
However, given the dense phase coverage obtained throughout the whole 
rotation cycle, the large-scale magnetic field of $\tau$~Sco (up 
to orders with $\ell\simeq 6$), as well as its non-variability on a time 
scale of about 1.5~yr, is very well constrained by our observations.

\section{Origin of the field}
\label{sec:disc}

Using the magnetic map we derived for $\tau$~Sco, and thanks in
particular to its unusual degree of complexity (by the
standards of early-B and O-type stars), several questions can be
addressed regarding the physics of massive stars.  Our results 
both give us
 the opportunity to rediscuss  the problem of the origin of magnetic
fields in very hot stars, and also enable us to investigate the impact
of complex fields on radiatively driven winds.

Although the classical picture was that magnetic fields of hot stars
were presumably fossil remnants from the formation stage, the
situation has changed considerably, with regular reports from both
observers and theoreticians that massive stars may be able to
generate dynamo processes, either
deep inside their convective cores \citep{Charbonneau01, Macdonald04,
Brun05}, within the greater part of their radiative envelope \citep{Spruit99,
Spruit02, Macdonald04, Mullan05, Maeder05, Braithwaite06}, or in a subsurface 
layer \citep{Tout95, Lignieres96}.  In each case, different processes are
invoked to explain the generation of magnetic fields.  How do these
proposals stand up in the particular case of $\tau$~Sco?

\subsection{Dynamo processes?}

Being intrinsically a very slow rotator (the second slowest rotator
among massive stars after HD~191612; \citealt{Donati06}), $\tau$~Sco
does not seem to be an optimal candidate for triggering dynamo
processes.  Of course, one may argue that the rotation rate of the
inner stellar regions may be far larger than that at the surface;
evolutionary models tend, however, to indicate that the radial gradient 
in rotation rate is only moderate in massive stars, particularly in
those hosting magnetic fields \citep{Maeder03, Maeder04, Maeder05}.

Wherever they operate (whether in the convective core, the radiative
envelope, or a subsurface layer), dynamo processes are all expected to
strengthen with rotation rate and to vanish when rotation is slow;
they should therefore be relatively weak in a star like $\tau$~Sco.  In the
particular case of the Spruit--Tayler dynamo 
processes \citep{Spruit99, Spruit02}, for example,
field strengths only of order a few G  are expected to appear at the
surface of a star with a rotation rate as small as that of $\tau$~Sco
\citep{Mullan05}, much lower than we have found.  In the case of the
core-dynamo hypothesis, \cite{Macdonald04} demonstrate that surface
magnetic fields need to originate in core fields that largely exceed
the equipartition value, which is again highly unlikely in a slowly
rotating star.

\begin{figure*}
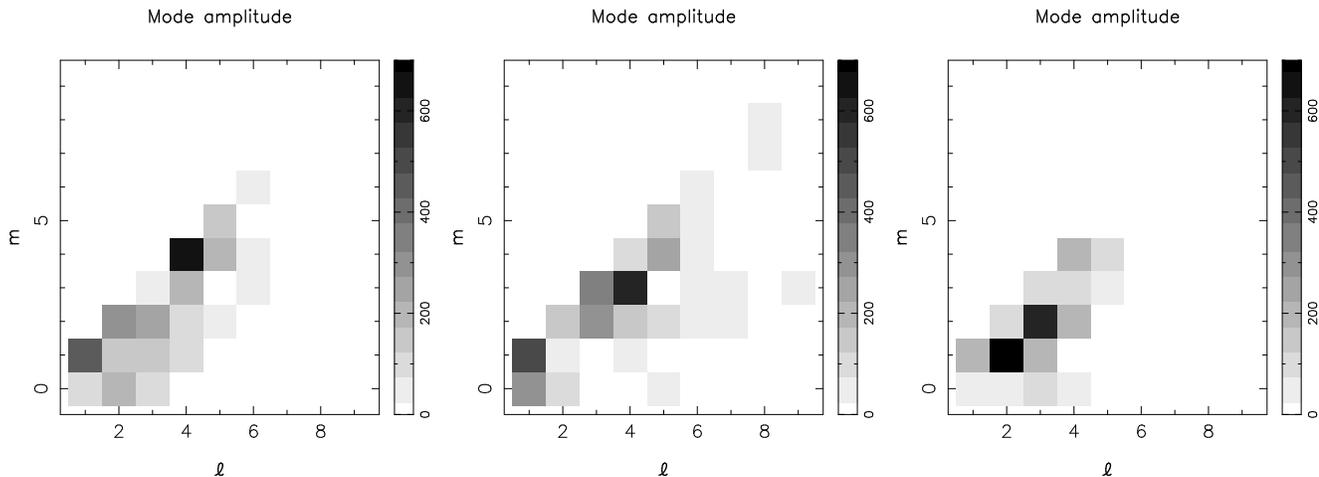

\center{\hbox{\includegraphics[scale=0.34,angle=-90]{fig/tausco_modes3.ps}\hspace{2mm}
              \includegraphics[scale=0.34,angle=-90]{fig/tausco_modes4.ps}\hspace{2mm}
              \includegraphics[scale=0.34,angle=-90]{fig/tausco_modes5.ps}}}
\caption[]{Modulus (in G) of the spherical-harmonic complex coefficients $\alpha_{\ell,m}$, 
$\beta_{\ell,m}$ and $\gamma_{\ell,m}$ (see Sec.~\ref{sec:mod}) for the reconstructed 
magnetic-field topology of $\tau$~Sco, as a function of mode degree $\ell$ and order $m$.   
These sets of coefficients respectively correspond to the magnetic images shown in 
the top panel of Fig.~\ref{fig:map} (radial field component), left panels of 
Fig.~\ref{fig:map2} (azimuthal and meridional component of potential field term) and 
right panels of Fig.~\ref{fig:map2} (azimuthal and meridional component of toroidal 
field term).  Only modes with $\ell<10$ are displayed here. }  
\label{fig:modes}
\end{figure*}

A second observation is that the magnetic features we reconstruct at
the surface of $\tau$~Sco are present at essentially all latitudes.
This is not compatible with the predictions of the Spruit--Tayler
dynamo model with buoyancy included to allow magnetic flux rise up to
the surface, which should produce magnetic
regions concentrated at intermediate latitudes \citep{Mullan05}.  Nor
is it compatible with the core-dynamo theory, which predicts that
flux tubes should mostly show up very close to the pole at the surface
of the star \citep{Macdonald04}.

Our observations further indicate that the reconstructed magnetic
field is mostly poloidal, and includes no more than a moderate surface
toroidal component.  This is not compatible with dynamos operating in
a shear layer below the surface or within the convective zone, which 
are expected to produce strong axisymmetric azimuthal fields 
\citep[e.g.,][]{Braithwaite06} that should likely show up at 
photo{\-}spheric level, and should even dominate the global magnetic 
map in the case of a sub-surface shear layer (as they do in
stars with very shallow subsurface convective zones; e.g.,
\citealt{Marsden06}).  Moreover, we observe that the photo{\-}sphere
of $\tau$~Sco experiences negligible latitudinal shear (at least 20
times smaller than that of the Sun).  Again, this is most probably
inconsistent with a subsurface shear-layer dynamo, which is expected
to generate azimuthal and radial gradients of angular velocity, at
least within the shear layer itself.

An additional argument against dynamo processes is that they should 
essentially succeed (if conceptually valid) at producing magnetic fields 
in most hot stars and not only in a small fraction of them.  The fact 
that magnetic fields are detected in a star like $\tau$~Sco, known for its 
peculiar spectroscopic morphology (e.g., through its abnormally strong UV 
P-Cygni lines and its unusually hard X-ray emission), after having been 
detected in other peculiar hot stars (like $\theta^1$~Ori~C, HD~191612 and 
$\beta$~Cep), represents further evidence that magnetic fields (at least 
those of moderate to high intensity) are not a common feature of most hot 
stars, but rather a rare occurence.  In this respect, the parallel between 
massive hot stars with intermediate-mass ones, for which only about 10\% 
of them are magnetic, seems strengthened by our new result.  

Finally, the magnetic configuration of $\tau$~Sco exhibits no sign of
variability evident on a timescale of a year or so.\footnote{We note
that two longitudinal-field estimates obtained for $\tau$~Sco by
\citet{Landstreet82}, $-61\pm30$~G and $-16\pm23$~G and on JDs
2,442,174.92 and 2,442,175.89 respectively, average to a longitudinal
field of $-33\pm18$~G at a phase 0.49 (cycle $-169$) according to
eqn.~\ref{eq:ephem}. The longitudinal field we measure at this phase,
$-20\pm3$~G, is compatible with the estimate given by
\citet{Landstreet82} to within the errors, and provides additional
(though admittedly fairly weak) evidence that the magnetic field of
$\tau$~Sco exhibits no intrinsic variability on timescales of
decades.}  Archival IUE data provide a further, albeit indirect,
indication that the field topology is probably stable on timescales of
decades.  Although all dynamo models proposed for hot stars predict
some kind of temporal variability, little information is available on
the typical timescale on which the surface fields are expected to
evolve.  At least for models depending on differential rotation, we
can expect the field configuration within the star to evolve on
timescales shorter than a year, as buoyant flux tubes typically need
timescales of order a year to travel to the surface in a star like
$\tau$~Sco \citep[provided the field in the flux tube is not too weak
compared to the local equipartition value;][]{Macdonald04}, again
implying that the observed non-variability argues against the proposed
dynamo models.

Recent numerical results by \citet{BraithSpruit04} and \citet{BraithNord06} 
also indicate that magnetic fields in hot stars seem to reach a stable 
equilibrium (involving both potential and toroidal fields) but are not 
self-amplified by instability processes, unless the star features self-sustained 
differential rotation \citep{Braithwaite06}.  Similar numerical results are 
obtained by Brun and collaborators (Zahn, personal communication).  Our result 
indicates that $\tau$~Sco is likely not a differential rotator (as it would 
otherwise host strong axisymmetric toroidal fields like those seen on cool 
stars, \citealt{Donati03});  it may therefore be fairly natural that no evidence 
for dynamo action is detected on $\tau$~Sco.  

\subsection{Fossil fields?}

The next step is to compare our observations with predictions of the
fossil-field theory.  Since $\tau$~Sco is rather young (a few Myr;
Sec.~\ref{sec:params}), the complexity of the field we detected is
probably not a problem; while low-order terms (with longer decay
times) are expected to dominate the fossil magnetic topologies of old
stars, higher-order terms should still be present in stars as young as
$\tau$~Sco.  Moreover, very little differential rotation and
variability (on a $\sim$yearly timescale) is expected to occur in
stars hosting superequipartition fossil magnetic fields, in agreement
with what we find.  

Both toroidal and poloidal fields of comparable strength are expected 
to be present within the star, at least to ensure dynamical stability 
of the fossil field on long timescales \citep{Moss01, BraithSpruit04, 
BraithNord06}.  The prediction is that the expected toroidal field 
should be roughly axisymmetric (with respect to the poloidal magnetic 
axis) and concentrate on the poloidal magnetic equator 
\citep{BraithSpruit04, BraithNord06}.  Although we indeed detect a small 
toroidal field at the surface of the star, its topology is not compatible 
with such predictions, which would require the toroidal field to coincide 
with the poloidal field equator (i.e.\ to show up mainly as a meridional 
field belt encircling the star, passing through both rotational poles and 
crossing the rotational equator at phases of about 0.3 and 0.8, see 
Sec.~\ref{sec:disc2}).  Note however that theory expects the toroidal 
field to remain within the stellar interior;  it may therefore be 
unsurprising not to detect it at photospheric level.  We suggest that 
the toroidal field structure we detect at photospheric level rather 
results from the interaction of the stellar wind and the magnetic 
field (see Sec.~\ref{sec:disc2}).  

Another attraction of the fossil-field theory is that there is no
contradiction with the fact that the star is both magnetic and slowly
rotating.  Actually, we note that magnetic hot stars are, in average,
even more slowly rotating than non-magnetic stars.  One can, of
course, wonder whether this is a real property of massive magnetic
stars, or simply an observational bias (magnetic fields being easier
to detect by spectro{\-}polarimetric methods in narrow-lined, slowly
rotating stars); however, experiments clearly demonstrate that several
rapidly rotating O stars (e.g., $\zeta$~Pup, $\zeta$~Ori) have surface
magnetic fields with strengths not more than a few tens of G (Donati,
in preparation).  The youth of $\tau$~Sco excludes the possibility
that slow rotation is a result of angular-momentum loss through a
magnetic wind during the main-sequence phase;\footnote{Following
\cite{Donati06}, we evaluate the magnetic-braking timescale of
$\tau$~Sco to be of order of 5~Gyr.  As this timescale is some 3 orders
of magnitude larger than the age of $\tau$~Sco, we can safely conclude
that angular-momentum loss through the current magnetically confined
wind is not responsible for the slow rotation.}  thus the most probable option is that
this situation is due to a process occuring during the formation
stage.  The idea proposed for magnetic, chemically-peculiar stars
(also more slowly rotating in average than non-magnetic stars of
similar spectral type), invoking magnetic coupling with a putative
accretion disc \citep{Stepien00}, is probably not applicable in the
case of massive stars, which do not exist as stars during the
formation stage and directly appear onto the main sequence.  One
possibility is that proto{\-}stellar discs with intrinsically higher
primordial magnetic fields are more successful at expelling angular
momentum from the disc (e.g., through magnetic jets) than those with
weak primordial fields, leading to magnetic hot stars rotating more
slowly than non-magnetic counterparts of similar mass.

\subsection{Conclusion}

Taking all these arguments into consideration, we find that the
magnetic topology we have reconstructed for $\tau$~Sco is more likely to be
of fossil origin than to be generated by any of the various dynamo
mechanisms proposed up to now in the literature.  If this is
confirmed, it would indicate that very hot magnetic stars probably
represent a high-mass extension of the classical Ap/Bp phenomenon; 
the reason these massive magnetic stars do not mark themselves
as chemically peculiar is probably related to their strong winds, which
prevent photo{\-}spheric element stratification building up.  This
scenario would also argue in favour of the proposition of
\citet{Ferrario05, Ferrario06}, who suggested that massive magnetic stars are the
progenitors of highly magnetic neutron stars.

\section{The extended magnetospheric structure}
\label{sec:disc2}

A second topic of interest is the impact of the magnetic
field on the radiatively driven wind of $\tau$~Sco.  In
particular, $\tau$~Sco gives us the opportunity of investigating
the confining effect of magnetic fields whose topology is more complex
than those of other early B and O stars for which similar studies have been
carried out \citep{Donati01, Donati02, Gagne05, Gagne05err}.

In the now-standard picture, initially proposed by \citet{Babel97} and
further investigated by \citet{Donati01, Donati02}, \citet{udDoula02},
\citet{Townsend05} and \cite{Gagne05, Gagne05err}, the magnetic field is assumed
to be dipolar; the dense wind coming from each magnetic hemisphere is
deflected by the field towards the magnetic equator, where it produces
a strong shock, a very hot X-ray emitting post-shock region (reaching
temperatures of $10^7$~K), and a cool, dense disk in the magnetic
equator, where the plasma accumulates before being ejected away from,
or accreted back onto, the star (depending on the local radial
velocity of the plasma when it reaches the disc, and on the effective
gravity in the disc at this point).

For a more complex magnetic topology, the picture is expected to
differ significantly.  The extended magnetic structure should show a
correspondingly greater degree of complexity, involving distinct
regions of closed loops alternating with regions of open field lines,
rather than two open-field polar cones and one closed-field magnetic
torus, as in the dipole field case.  Wind flows should freely escape
the star along open field regions and should produce shocks and very
hot X-ray emitting plasma within each closed-field region, with cool,
dense condensations forming at loop summits.  The resulting
magneto{\-}spheric structure should therefore begin to resemble that of
the Sun, with hot coronal arcades confining cool, dense,
prominence-like structures (the main difference being, of course, the
heating mechanism itself).  

\begin{figure*}
\center{\hbox{\includegraphics[scale=0.4]{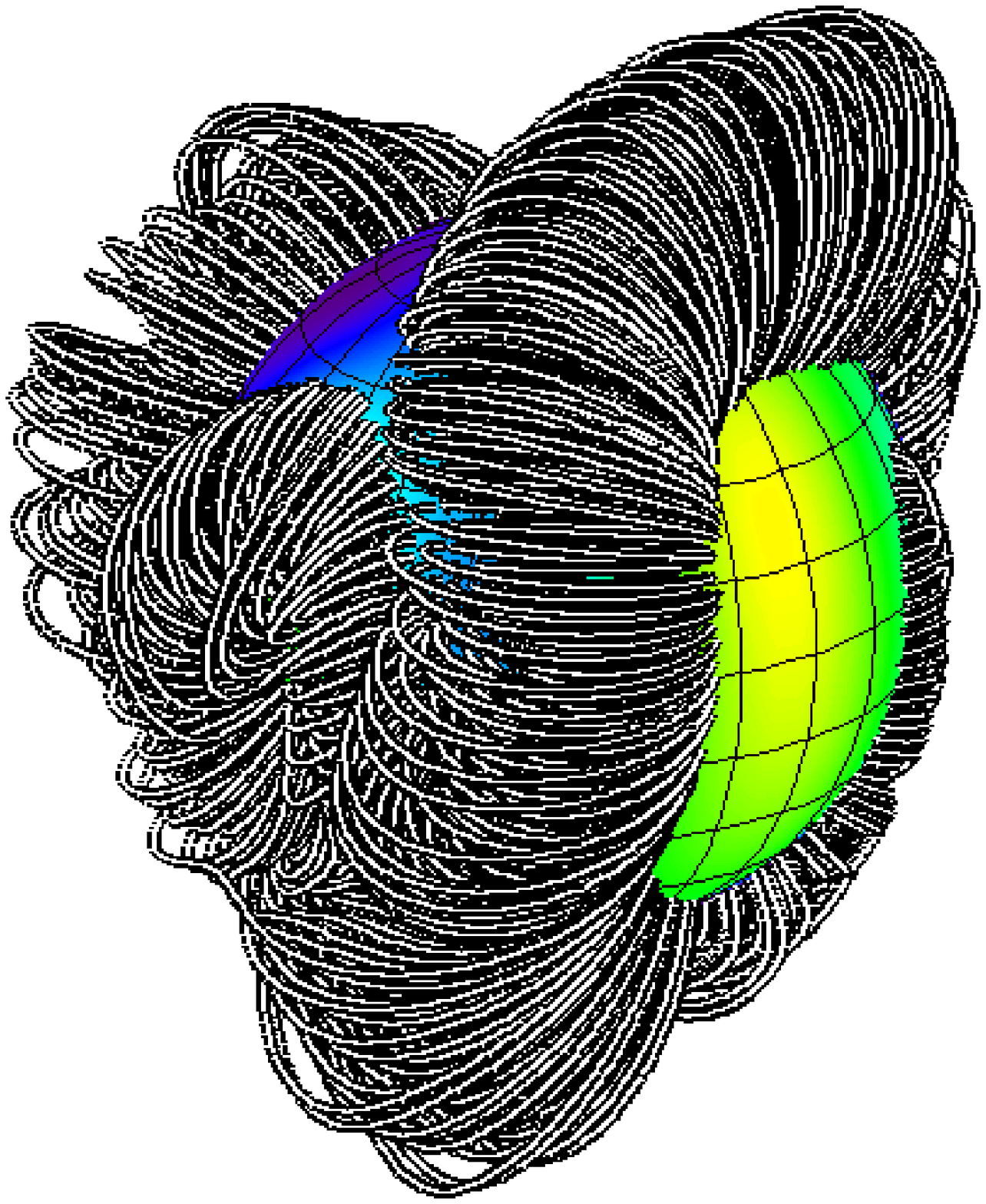}
              \includegraphics[scale=0.4]{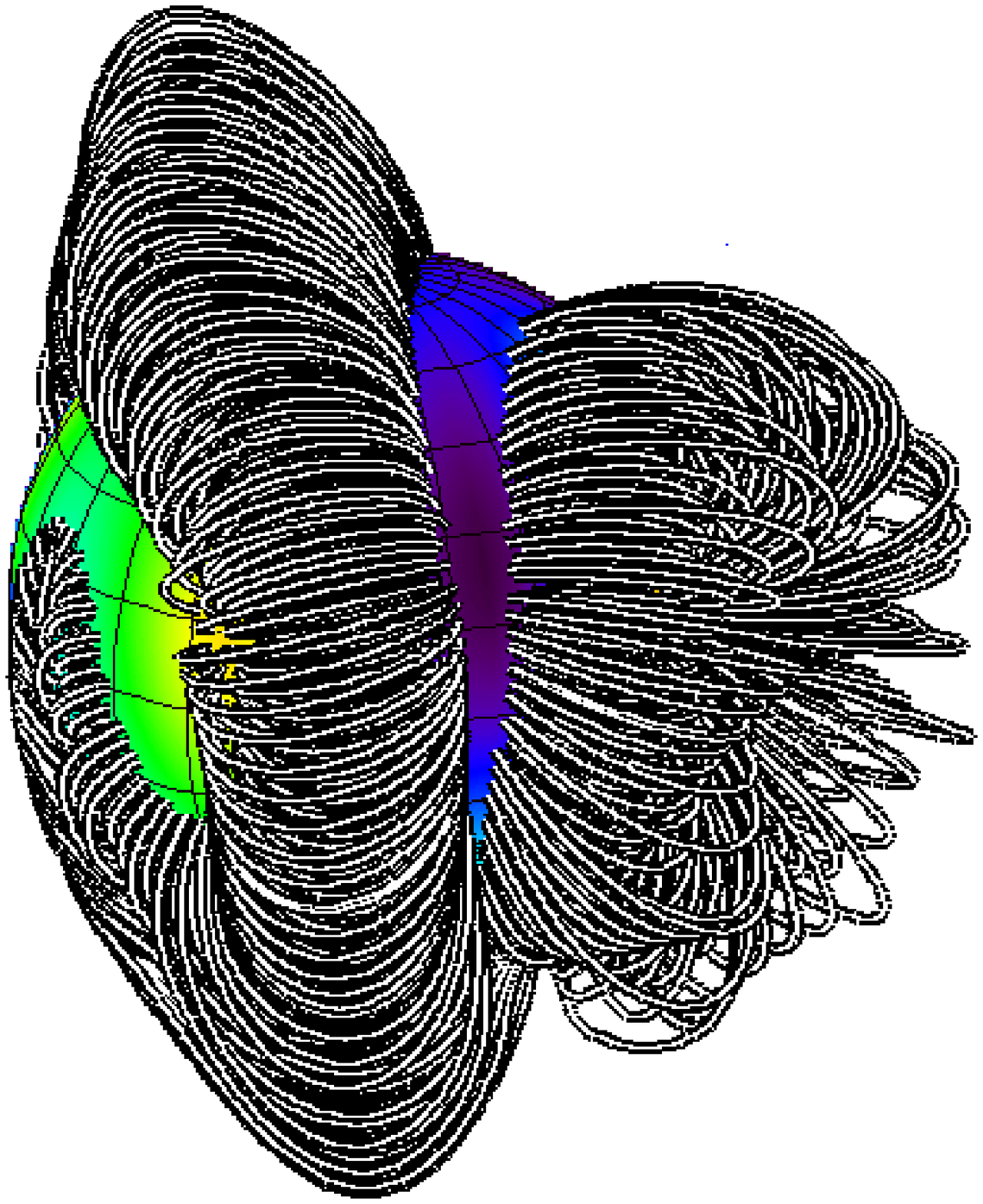}}} 
\caption[]{Closed magnetic-field lines of the extended magnetic
configuration of $\tau$~Sco, extrapolated from the photo{\-}spheric
map of Fig.~\ref{fig:map}.  The star is shown at phases 0.25 (left) 
and 0.83 (right).  Note the warp of the magnetic equator and the 
additional network closed loops around phase 0.65 (mostly visible 
on the right side of the right panel).} 
\label{fig:cor}
\end{figure*}

\subsection{The magnetic-confinement parameter}

First, to evaluate to what distance from the star the wind of $\tau$~Sco is
magnetically confined, it is useful to consider the wind
magnetic-confinement parameter $\eta$, defined by \citet{udDoula02}
to characterize the ratio between the magnetic-field energy density and
the kinetic energy density of the wind:
\begin{equation}
\eta = B^2 \rstar^2 / \Mdot \vinf,
\end{equation}
\noindent where $B$ is the typical magnetic-field strength, \Mdot\ is 
the average
mass-loss rate, and \vinf\ is the terminal wind velocity.  Our study
demonstrates that $B$ is in average $\simeq300$~G over the surface of 
$\tau$~Sco.  However, there is a
substantial dispersion in published
estimates of \Mdot, with recent determinations ranging 
$0.2\mbox{--}6\times10^{-8}$~\mspy\ \citep[e.g.,][]{Mokiem05, Repolust05}; we
adopt a value of $2\times10^{-8}$~\mspy\ as a reasonable average
of modern observational determinations.  
Observations also suggest a terminal velocity of about 2,000~\kms\
\citep[e.g.,][]{Abbott78}, but line-driven wind theories predict a
significantly higher value (up to 3,800~\kms,
\citealt{Pauldrach87}; note, however, that the stellar parameters used
for these theoretical studies, and in particular the mass,
temperature, luminosity and radius of $\tau$~Sco, are all
significantly overestimated compared to modern values). We adopt
$\vinf = 2,000$~\kms\ \citep{Mokiem05}.  

With these values, we find that the wind confinement parameter is about
40, and thus that the Alfven radius, above which all closed magnetic
loops open under the wind ram pressure, is of order of 2~\rstar.  This
is in good agreement with the findings of \citet{Cohen03}, based on
Chandra observations (and in particular on \sixiii\ and \mgxi\ line ratios), 
that the X-ray emitting plasma is located at average distances of about 
1~\rstar\ above the surface\footnote{A recent re-analysis of the Chandra 
$\tau$~Sco spectra using updated atomic data \citep[e.g.,][]{Gagne05err} 
indicates that all line ratios from He-like ions (including both \sixiii\ 
and \mgxi) are consistent with the X-ray emitting plasma being concentrated at a 
distance of about 1~\rstar\ above the photosphere (Cohen 2006, personal
communication).}.

Arguably, this agreement may be partly coincidental, given
the large uncertainty on \Mdot;  if it were 10 times smaller
than the value we adopted, with all other parameters held fixed,
$\eta$ would reach a value of 400. The magneto{\-}sphere would then be
confined out to significantly larger distances (typically of order
4\rstar); we should then detect hard X-ray emission from loops
extending several \rstar\ above the stellar surface, which is
apparently not the case.  If \Mdot\ were larger than
$6\times10^{-8}$~\mspy, then $\eta$ would be smaller than 10, and no
stable magnetic loops extending further than 0.3~\rstar\ above the
surface would survive the wind pressure \citep[e.g.,][]{udDoula02},
again in contradiction with X-ray observations.  We therefore conclude
that, if this model is correct, X-ray observations constrain $\eta$ to
values ranging typically within 20 to 100, and 
hence the mass-loss rate to values of 
1--4$\times10^{-8}$~\mspy.

\begin{figure}
\center{\includegraphics[scale=0.4]{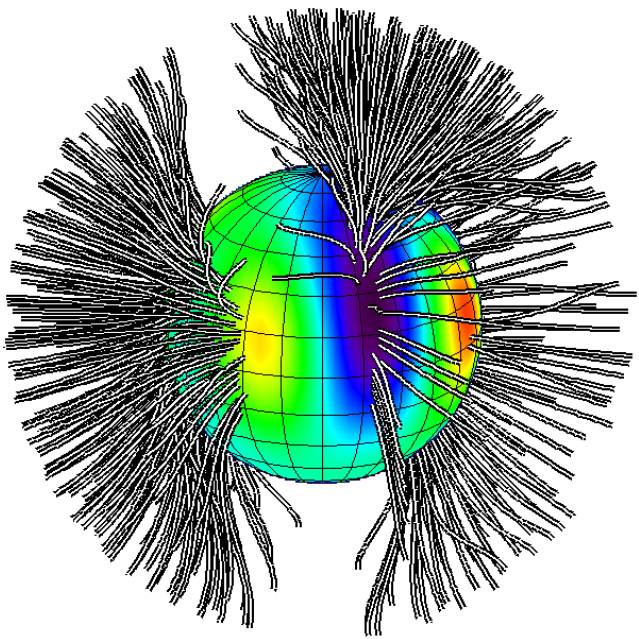}} 
\caption[]{Open field lines of the extrapolated extended magnetic
configuration.  The star is shown at phase 0.83 only. }
\label{fig:cor2}
\end{figure}

\subsection{Magnetic-field extrapolation}

Using the field-extrapolation technique of \citet{Jardine99}, with a
spherical source function set to 2~\rstar\ (to mimic the mainly radial
orientation of the field lines at distances larger than 2~\rstar), we
can draw inferences on the large-scale magneto{\-}spheric structure of
$\tau$~Sco, presented in Figs.~\ref{fig:cor} (closed field lines) and 
\ref{fig:cor2} (open field lines).
We find that the extended field structure is significantly more complex 
than a global dipole, even though it still features a torus of closed 
magnetic loops encircling the star (with an axis of symmetry roughly 
tilted at $\simeq$90$\deg$ to the rotation axis; Fig.~\ref{fig:cor}) 
and two main cones of open field lines on opposite sides of 
the star (see Fig.~\ref{fig:cor2}).  In particular, the magnetospheric 
equator is significantly warped, and additional networks of closed loops are 
present at low latitudes (e.g., one around phase 0.65, left panel of 
Fig.~\ref{fig:cor}, in conjunction with the equatorial region of positive 
radial field reconstructed at this phase, another one around phase 0.4). 
Most closed loops typically extend to a distance of up to 2~\rstar, in
reasonable agreement with constraints derived from Chandra data.

Note that these small networks of closed loops (at phases 0.40 and 0.65) 
both coincide with local maxima of the reconstructed toroidal field 
component (see top right panel of Fig.~\ref{fig:map2});  
from this apparent spatial correlation, we speculate that 
the toroidal field component we detect may be produced through 
an interaction of the stellar wind and the magnetic field right at 
the photospheric level.  
We also note that these loops roughly coincide with the rotation phase 
at which the unpolarised line profile of $\tau$~Sco is slightly narrower 
than average;  again, this may reflect the particular wind configuration 
that results from this specific field configuration at the stellar 
surface.  MHD simulations are of course needed to confirm whether this 
idea is realistic or not.  


\subsection{Observational implications}
\label{sec:disc3}

\subsubsection{Optical and UV diagnostics}
\label{sec:disc3a}

This model implies that excess absorption in UV lines should
occur when the magneto{\-}spheric equatorial plane crosses the line of
sight, as in all similar hot magnetic stars
\citep[e.g.,][]{Donati01, Donati02, Donati06, Neiner03}, i.e., at
phases 0.3 and 0.8 -- in excellent agreement with observations (see
Fig.~\ref{fig:UVfig3}).  In particular, the fact that the two
UV absorption events we detect are largely similar in shape, and separated by
0.5~rotation cycles, provides independent confirmation that either the angle 
$i$ of the rotation axis to the line-of-sight (which we found to be
$\simeq70\degr$), or the global tilt of the 
large-scale magnetic structure to the rotation axis (for which we derived an 
estimate of $\sim90\degr$), or both, is/are large.  
As mentioned previously, the relative phasing between
our new spectro{\-}polarimetric data and the old archival IUE spectra is
accurate enough (of order 1\%; Sec.~\ref{sec:rotIUE}) to ensure that
this match is not a coincidence; it thus provides a strong argument in
favour of the present model.

Photo{\-}metric measurements secured by Hipparcos indicate a constant
flux level (to within 10~mmag), showing that the column density of the
wind material trapped within the magneto{\-}spheric equator of $\tau$~Sco
is not high enough to produce detectable light variations through
scattering, even when the disc is seen edge-on.  This situation is
similar to that found for $\beta$~Cep and $\theta$~Ori~C (for
which no photo{\-}metric variations are detected; \citealt{Donati01,
Donati02}), but differs from HD~191612 and $\sigma$~Ori~E (for
which eclipses of the continuum radiation by the magneto{\-}spheric plasma are
observed, at levels of 0.04 and 0.15~mag, respectively;
\citealt{Walborn04, Townsend05,  Donati06}).  

In our data,
we detect no rotational modulation of the H$\alpha$ flux from
$\tau$~Sco, mimicking what is observed in $\beta$~Cep (which shows
only long-term H$\alpha$ variations), but very different to
$\theta$~Ori~C, $\sigma$~Ori~E and HD~191612 (all of which exhibit
strong H$\alpha$ modulation; \citealt{Stahl96, 
Townsend05, Walborn03}).  More modelling is required to check whether these
observations are compatible with the basic picture presented here;
this is postponed for a future study.

\subsubsection{X-ray diagnostics}

The wind pressure at the base of the the postshock region is given by
\begin{equation}
p_{\rm w} = \epsilon\ \Mdot \vinf / 4 \pi \rstar^2 ,
\end{equation}
to first order,
where $\epsilon = 1/x^2 - 1/x^3$ and $x$ is the radial distance 
from the centre of the star at which the equilibrium location of 
the shock front settles  (in units of $\rstar$).  For a range of 
reasonable values of $x$ ($\sim$1.2--1.7), $\epsilon$ remains 
roughly constant ($\sim$0.10--0.15), implying a wind pressure 
of about 20--30~\gpcps.  This corresponds to a number 
density of protons and electrons of about $10^{10}$~\pcc\ for a 
postshock temperature of order $10^7$~K within
the loop.  Again, this is in good agreement with the
upper limits on the electron density derived by \citet{Cohen03} from
Chandra data.  By assuming that the closed `corona' (the equatorial
magneto{\-}spheric torus and the small additional networks of closed 
loops) is filled with such a plasma, and using the simple
coronal-structure model of \citet{Jardine02}, we find that the
resulting emission measure is of order of a few $10^{54}$~\pcc, in
reasonable agreement with actual measurements from X-ray spectra
\citep{Wojdowski05}.  This result essentially indicates that, in the
model we have devised, the observed X-ray emission can be mostly attributed
to the magneto{\-}sphere (and that the model is therefore broadly
consistent with X-ray observations).


If our speculation is correct, it implies that the X-ray emission of
$\tau$~Sco should be modulated on a timescale equal to the rotation
period (i.e., 41~d), as a result of the magneto{\-}sphere being partially
eclipsed by the stellar disc in appropriate viewing configurations.
We estimate the expected fractional modulation to be about
40\%, i.e., comparable to that of $\theta$~Ori~C
\citep{Gagne05err, Gagne05}, but should feature two main eclipse episodes each
rotation cycle, centred on phases $\sim$0.3 and 0.8 (concomitant with
the UV line-absorption events witnessed in IUE spectra);  this should
be easily detectable given adequate temporal sampling.  

Checking such predictions should provide a strong test of our model.
If the observed X-ray rotational modulation is much weaker than
expected, it could imply that most of the magneto{\-}spheric emission is
produced in small-scale loops evenly spread over the stellar surface;
this would argue for additional, high-order components of the
magnetic topology that we are not able to detect in this study (as a
result of the limited spatial resolution provided by Doppler imaging
for stars rotating as slowly as $\tau$~Sco).  

At this stage, a more accurate model is obviously necessary to confirm
the conclusions of the present paper, and to develop in more detail
how well the X-ray spectrum of $\tau$~Sco 
can be reproduced once the
magneto{\-}spheric structure, including the wind-induced expansion of
magnetic loops, is consistently taken into account.  Such a
sophisticated model should also aim to reproducing the observed
modulation of UV spectra and the upper limits on the photo{\-}metric
and H$\alpha$ flux variability (e.g., as done for $\sigma$~Ori~E by
\citealt{Townsend05b}).

\section{Conclusion}

We have reported the detection of a magnetic field on
the massive B0.2$\;$V star $\tau$~Sco, using data obtained mostly with
ESPaDOnS, the new high-resolution stellar spectro{\-}polarimeter
recently installed at CFHT.  From the Zeeman signatures and their
temporal variability, we were able to identify the rotation period of
$\tau$~Sco and to reconstruct the large-scale topology of its
photo{\-}spheric field.  Archival IUE spectra confirm that the
rotational modulation is stable on timescales of decades, with $P_{\rm
rot} \simeq 41.03$~d.

We find that the surface magnetic topology is unusually complex
(judged by the small sample of massive-star results) and is mostly
potential.  It also includes a moderate toroidal component; in particular,
the strength of this toroidal component (relative to 
that of the poloidal component) is much lower than that found
in partly-convective cool stars hosting dynamo-generated magnetic fields.  No
temporal variability of the magnetic structure is detected over the
1.5-yr period of our observations; we thus conclude that any surface
differential rotation of $\tau$~Sco is at least 20 times weaker than
that of the Sun.

We determine that the large-scale magneto{\-}spheric structure of $\tau$~Sco
is significantly more complex than a global dipole;  it features in particular 
a significantly warped torus of closed magnetic loops encircling the star,  
tilted at about 90\degr\ to the rotation axis, as well as additional  
(smaller) networks of closed field lines.  
The extended magnetic topology we derive from extrapolations of
the photo{\-}spheric magnetic maps is apparently compatible with the
published X-ray luminosity and spectral characteristics of $\tau$~Sco.
Our model is also compatible with the observed modulation of UV
spectral lines.  We predict that $\tau$~Sco should
exhibit a clear rotational modulation of its X-ray emission.

From these results, we conclude that its magnetic
field is most probably a fossil remnant from the formation stage.  
We cannot yet completely rule out the possiblity that the field is 
produced through one of the recently-elaborated dynamo processes that may 
operate in the radiative zones of hot stars, but our findings already 
indicate that this option is rather unlikely.

\section*{Acknowledgements}

We thank the CFHT staff for their help during the various runs with ESPaDOnS.  
We also thank the referee, R.~Townsend, as well as D.~Cohen, N.~Walborn and 
M.~Smith for valuable suggestions and comments that improved the manuscript.


\bibliography{tausco}

\bibliographystyle{mn2e}

\end{document}